\documentclass[a4paper,11pt]{article}
\usepackage{jheppub} 
\usepackage{lineno}
\usepackage{graphicx}
\usepackage{lipsum}
\usepackage{slashed}
\usepackage{amsmath}
\usepackage[normalem]{ulem}
\usepackage{enumitem}\usepackage{mathtools}
\usepackage{bm}
\usepackage[all]{hypcap}
\usepackage{tikz}
\usepackage{tikz-feynhand}
\usepackage{bbm}
\usepackage{xcolor}
\usepackage{subcaption}
\usepackage[font=footnotesize,labelfont=bf]{caption}
\usepackage{mathabx}
\usepackage{pifont}
\usepackage{multirow}
\usepackage{tablefootnote}
\usepackage{cleveref}
\usepackage{physics}

\newcommand{\cC}{\mathcal{C}}
\newcommand{\Op}{\mathcal{O}}

\title{
The Price of a Large Electron Yukawa Modification
}

\author[a]{Lukas Allwicher,} 
\emailAdd{lukas.allwicher@desy.de}

\author[b]{Matthew McCullough,} 
\emailAdd{matthew.mccullough@cern.ch}

\author[c]{Sophie Renner,}
\emailAdd{sophie.renner@glasgow.ac.uk}

\author[b,d,e]{Duncan Rocha}
\emailAdd{drocha@uchicago.edu}

\author[c]{and Benjamin Smith}
\emailAdd{b.smith.4@research.gla.ac.uk}

\affiliation[a]{Deutsches Elektronen-Synchrotron DESY, Notkestr. 85, 22607 Hamburg, Germany}
\affiliation[b]{Theoretical Physics Department, CERN, 1211 Geneva, Switzerland}
\affiliation[c]{School of Physics and Astronomy, University of Glasgow, Glasgow G12 8QQ, United Kingdom}
\affiliation[d]{Theoretical Physics Division, Fermi National Accelerator Laboratory, Batavia, IL 60510}
\affiliation[e]{Enrico Fermi Institute, Physics Department, University of Chicago, Chicago, IL 60637}

\abstract{The theoretical implications of an electron Yukawa modification are considered in the context of a possible Higgs pole run at FCC-ee, aimed at bounding this coupling. We start from an effective field theory viewpoint, considering the impact of renormalisation group effects on related observables and also examining assumptions on the broader UV flavour structure.  We then give an overview of the landscape of simplified models, investigating phenomenological constraints arising at higher orders.  A short discussion of fine-tuning is also included.}

\begin{document}
\preprint{CERN-TH-2025-201, DESY-25-142}

\maketitle
\section{Introduction}
\label{sec:intro}

Within the Standard Model (SM), the masses of the fundamental fermions are in direct correspondence with their Yukawa couplings to the physical Higgs boson. Given that we know the electron mass very accurately, an improved bound on the electron Yukawa coupling would not tell us anything new about the SM itself, rather the insight gained would be entirely about physics beyond the SM, which can break this degeneracy between the mass and coupling. Current constraints on the Higgs-electron coupling are very weak, at about 240 times the SM value~\cite{CMS:2022urr}, while HL-LHC projections can reduce this to about 120 times the SM value~\cite{Cepeda:2019klc}. A proposed Higgs-pole run at FCC-ee could bring these constraints down to $O(1)$ of the SM value~\cite{Greco:2016izi,dEnterria:2021xij,FCC:2025lpp} (or even reach a measurement of the SM if transverse polarization of the beams can be achieved without sacrificing luminosity~\cite{Boughezal:2024yjk}). However this is very challenging experimentally, requiring about a year of runtime and highly monochromatised beams \cite{Jadach:2015cwa,Telnov:2020rxp,Faus-Golfe:2021udx,Zhang:2024sao,Blondel:2025kjz,Zhang:2025wxn}. Given these challenges, it is worth assessing the full space of beyond the Standard Model (BSM) theories that could be tested by this measurement, asking which models would be first seen here, how theoretically plausible these models are, and whether future measurements of other observables could test the same theory space more easily. There are many ways of defining `plausibility' of a model, all of which are subjective. We explore a few different criteria, namely: how aligned and electron-specific does the flavour structure need to be? Is there a large hierarchy of couplings needed in any particular model? How fine-tuned and radiatively stable would such a model be?

We begin with a general effective field theory (EFT) approach in Sec.~\ref{sec:EFT}. 
There is just one effective operator in the Standard Model EFT (SMEFT) Warsaw basis~\cite{Grzadkowski:2010es} which can contribute to the electron Yukawa by an amount which is unsuppressed by its SM value, and without contributing to other fermion Yukawas:
\begin{equation}
    \mathcal{O}_{eH} = |H|^2 H \overline{l} e.
\label{eq:Yuk6}
\end{equation}
Since direct bounds allow a very significant enhancement of the electron Yukawa, one can hope to find indirect constraints on this operator from its renormalisation group (RG) flow into other operators which contribute to better measured observables.
Interestingly, from an RG perspective, $\mathcal{O}_{eH}$ is relatively isolated due to helicity non-renormalisation theorems~\cite{Cheung:2015aba} and approximate flavour symmetry in the anomalous dimensions, thus it generates no other phenomenologically relevant operators at one loop. At two loops, electroweak dipole operators are generated~\cite{Panico:2018hal,EliasMiro:2020tdv, Brod:2022bww, Davila:2025goc}, and the magnetic dipole moment of the electron can provide indirect constraints on very large modified electron Yukawas.  We also consider the possibility of radiatively generating $\mathcal{O}_{eH}$ in the IR as a result of the existence of some other non-vanishing Wilson coefficient, finding that for this scenario complementary signatures may appear at HL-LHC and FCC-ee, but that nevertheless some operators involving top or bottom quarks could first make themselves known in a modification of the electron Yukawa.

If the new physics which generates the operator~\eqref{eq:Yuk6} couples to different generations of leptons, then there may be modifications to the Yukawa couplings of the muon and the tau as well as lepton flavour violating signatures. These are much better constrained than the electron Yukawa, and we examine the current and future sensitivities on the flavour structure of any such new physics.

In the second part of this work we focus on explicit classes of perturbative UV models generating $\mathcal{O}_{eH}$ at tree and one-loop level, finding that generically the phenomenological constraints on concrete scenarios are stronger than would be expected based on a purely EFT approach. One reason for this is that in realistic scenarios, integrating out a heavy field generates a set of operators rather than $\mathcal{O}_{eH}$ alone. In particular, this can occur at lower perturbative order than generating SMEFT operators radiatively from $\mathcal{O}_{eH}$, leading to enhanced experimental signals.
The connection between modified Yukawas and the anomalous magnetic moment of the electron is therefore stronger in nearly all models of new physics than it is within the EFT. The only simple exceptions to this are models with an additional Higgs doublet, where the dipole operators arise at the same perturbative order when integrating out the heavy Higgs as they do from RG flow.
For this reason, models with an additional Higgs doublet persist as the least constrained scenario.  General comments on fine-tuning are provided before summarising.

This is far from the first paper to consider significantly modified leptonic Yukawa couplings.  Early studies focussed on answering a broader flavour question; namely the hierarchies present in the fermion masses, which could potentially be explained if the lighter generation masses arose at higher EFT orders in the presence of a scale separation \cite{Babu:1999me,Giudice:2008uua}.  
Later works have explored phenomenological and model-building aspects of Higgs-lepton coupling deviations (e.g.~\cite{Craig:2012vn,Dery:2013rta,Altmannshofer:2015qra,Altmannshofer:2016zrn,Chen:2017nxp,Dery:2017axi,Botella:2018gzy,Han:2021lnp,Alonso-Gonzalez:2021tpo}), including the possibility of lepton flavour violating decays (e.g.~\cite{Goudelis:2011un,Davidson:2012ds,Dery:2014kxa,Aloni:2015wvn,Altmannshofer:2015esa,Banerjee:2016foh,Hayreter:2016aex,Buschmann:2016uzg,Zhang:2015csm,Chang:2016ave,Belusca-Maito:2016axk,Huitu:2016pwk,Barman:2022iwj,Abu-Ajamieh:2025jsz}), and connections to anomalous magnetic moments (e.g.~\cite{Chen:2015vqy,Wang:2016rvz,Fajfer:2021cxa,Davoudiasl:2023huk}).
More recently, model building studies for the electron Yukawa have been inspired by the possible FCC-ee prospects~\cite{Chang:2022eft,Davoudiasl:2023huk,Erdelyi:2025axy}. Related work has also been done on CP violating Yukawa couplings in the lepton sector, and their connection to electric dipole moments or baryogenesis (e.g.~\cite{King:2015oxa,Egana-Ugrinovic:2018fpy,DeVries:2018aul,Fuchs:2019ore,Fuchs:2020uoc,Cheung:2020ugr,Altmannshofer:2020shb,AharonyShapira:2021ize,Bahl:2022yrs,Brod:2022bww,Kosnik:2025srw,Nhi:2025iob}).

Our study extends and is complementary to these works, in particular concerning possible classes of UV flavour structure, the impact of two-loop RG, an extended discussion of fine-tuning, and by considering current and future constraints on operators and models which can contribute to $\Delta \kappa_e$ at the one-loop level.

\section{Enhancing the electron Yukawa in SMEFT}
\label{sec:EFT}
At the electroweak (EW) scale, the effects of decoupled new physics are well described by higher-dimensional operators in the SMEFT framework. In the Warsaw basis \cite{Grzadkowski:2010es}, there are several operators which can shift the electron Yukawa, however we focus on $\mathcal O_{eH}$, defined in Eq.~\eqref{eq:Yuk6}, as it allows for a non-universal Yukawa coupling modification.
Thus we consider the following Lagrangian for the electron Yukawa coupling up to dimension six:
\begin{equation}
    -\mathcal{L}_{\text{Yuk}} = y_e H \bar l e  - \cC_{eH} \underbrace{|H|^2 H \bar l e}_{\mathcal{O}_{eH}} +... ~~,
    \label{eq:effop}
\end{equation}
where $l$ and $e$ are here the first generation left-handed lepton doublet and right-handed lepton singlet respectively. 

Importantly, this new dimension-6 term modifies the relationship between fermion mass and Higgs coupling from the SM prediction. Assuming no operators of dimension higher than those in Eq.~\eqref{eq:effop} are important, we may proceed to calculate the ratio of the Yukawa coupling to its SM value:
\begin{equation}
    \kappa_e = 1-\frac{v^3}{\sqrt{2}\,m_e} \cC_{eH}.
\end{equation}
Here and throughout, we assume that the $\cC_{eH}$ coefficient is purely real. Imaginary parts have strong indirect constraints from electric dipole moments \cite{Brod:2022bww,Panico:2018hal}, and are not necessary to generate the Higgs coupling deviations that we focus on.
In the SM, the electron Yukawa $y_e$ is very small, of $\mathcal O(10^{-6})$ in natural units. It appears that the effects of new physics could easily dominate such a small value, even with a moderate or large separation between $v$ and the scale of new physics $\Lambda$. However, understanding both the fine-tuning and RG flow of such a theory leads to theoretical insights on the parameter space. This will be discussed in Sec.~\ref{sec:radiative}.  Remaining within the context of the EFT description, in this section we investigate the general implications of the $\mathcal{O}_{eH}$ operator; specifically, how it can be connected to other operators via RG flow, and the model-independent constraints or future tests this can enable for $\kappa_e$.
We also consider various assumptions concerning the EFT
flavour structure, with varying degrees of plausibility and motivation, and how they may be constrained by complementary experimental probes. 
In Sec.~\ref{sec:models} we will go on to consider explicit models under certain flavour structure assumptions and the additional EFT operators they give rise to.

\subsection{Loop level connections to other SMEFT operators}
\label{sec:EFTdipoles}
The operator $\mathcal O_{eH}$ will mix with other dimension-6 SMEFT operators under RG, producing a set of operators emergent in the IR. Spurion analysis allows us to quickly determine the operators in this set by considering the flavour symmetries of the SM. Under the global $U(3)^5$ flavour group, the $\Op_{eH}$ operator is charged as:
\begin{equation}
    \Op_{eH} \sim \bar{\mathbf{3}}_L \times \mathbf{3}_e \,.
\end{equation}
In general, it can mix under RG with other operators with the same flavour charge, or with different flavour charges as long as corresponding spurionic factors of SM Yukawas enter the running. This immediately restricts the number of dimension-6 SMEFT operators which can be connected to $\Op_{eH}$ via renormalisation group equations (RGEs), at any loop order. Specifically, the complete list of operators with the same flavour charge as $\Op_{eH}$ consists only of the leptonic electroweak dipoles:
\begin{align}
\Op_{eB} &= (\bar \ell {\sigma _{\mu \nu }}e)H\, B^{\mu \nu } \sim \bar{\mathbf{3}}_L \times \mathbf{3}_e, \label{eq:eBop}\\
\Op_{eW} &= (\bar \ell {\sigma _{\mu \nu }}e){\tau ^I}H\, W_I^{\mu \nu } \sim \bar{\mathbf{3}}_L \times \mathbf{3}_e. \label{eq:eWop}
\end{align}
An additional four operators carry the same charge under $U(3)_L\times U(3)_e$ as $\mathcal{O}_{eH}$, whilst also carrying additional SM flavour charges:
\begin{align}
&\Op_{lequ}^{(1)} = (\bar \ell^\alpha e)
\epsilon_{\alpha\beta} (\bar q^\beta u)\sim \bar{\mathbf{3}}_L \times \mathbf{3}_e \times \bar{\mathbf{3}}_Q \times \mathbf{3}_u,\\
&\Op_{lequ}^{(3)} = (\bar \ell^\alpha\sigma_{\mu\nu} e)
\epsilon_{\alpha\beta} (\bar q^\beta \sigma^{\mu\nu}u)\sim \bar{\mathbf{3}}_L \times \mathbf{3}_e \times \bar{\mathbf{3}}_Q \times \mathbf{3}_u,\\
&\Op_{ledq} = (\bar \ell^\alpha e)(\bar d q^{\alpha})\sim \bar{\mathbf{3}}_L \times \mathbf{3}_e \times \bar{\mathbf{3}}_d \times \mathbf{3}_Q,\\
&\Op_{le} = (\bar \ell \gamma^\mu \ell)(\bar e \gamma^\mu e)\sim \bar{\mathbf{3}}_L \times \mathbf{3}_e \times \bar{\mathbf{3}}_e \times \mathbf{3}_L.
\end{align}
The SM Yukawas $Y_u$, $Y_d$ and $Y_e$, can be thought of as having the spurionic flavour charges $Y_u\sim \mathbf{3}_Q \times \bar{\mathbf{3}}_u$, $Y_d\sim \mathbf{3}_Q \times \bar{\mathbf{3}}_d$ and $Y_e\sim \mathbf{3}_L \times \bar{\mathbf{3}}_e$ respectively, meaning that each of the four operators above has the same flavour charge as one of $Y_u^\dagger\, \Op_{eH}$, $Y_d\, \Op_{eH}$ or $Y_e\, \Op_{eH}$. If the operators involve first generation lepton indices together with third generation quark or lepton flavour indices, they can be connected via running to $\Op_{eH}$ by amounts proportional to the large top Yukawa or (much smaller) bottom or tau Yukawas.

The isolation of $\Op_{eH}$ is exacerbated further by helicity non-renormalisation theorems~\cite{Cheung:2015aba}, which disallow its running at one loop into any operator other than the six-Higgs operator $\Op_{H}$ (for which running is anyway suppressed by tiny lepton Yukawas, by the flavour arguments above).
\cite{Panico:2018hal}, gives the RGE
The leading RGEs describing the running from $\Op_{eH}$ into the EW dipoles therefore arise at the two-loop level \cite{Panico:2018hal, EliasMiro:2020tdv, Brod:2022bww, Davila:2025goc}:\footnote{The overall sign of this RGE is inconsistent between Ref.~\cite{Davila:2025goc} and Refs.~\cite{Panico:2018hal, EliasMiro:2020tdv, Brod:2022bww}. We use the sign from the latter references.}
\begin{equation}
\label{eq:eftdipoleRGE}
\frac{d}{d\ln \mu}
\left(\begin{array}{c} \cC_{eB}\\ \cC_{eW}\end{array}\right)
=\frac{g^3}{(16\pi^2)^2}\frac{3}{4}
\left(
\begin{array}{c}
\tan\theta_W \mathcal{Y}_H + 4 \tan^3\theta_W \mathcal{Y}_H^2 (\mathcal{Y}_L +\mathcal{Y}_e)
\\
\frac{1}{2} + \frac{2}{3} \tan^2\theta_W \mathcal{Y}_H (\mathcal{Y}_L+\mathcal{Y}_e)
\end{array}
\right)
\cC_{eH}\\,
\end{equation}
where $\theta_W$ is the Weinberg angle and $\mathcal{Y}_X$ is the hypercharge of the field $X$.
The covariant derivative we use is defined $D_\mu=\partial_\mu+ig \frac{\tau^I}{2}W_\mu^I+ig^\prime \mathcal{Y}B_\mu$.
Since some of the phenomenology of the leptonic dipoles is measured very precisely (e.g. $\Delta a_e$, $\mathcal{B}(\mu\to e \gamma)$), it is worth investigating whether this mixing, even at the two loop level, can provide relevant constraints on the coefficient of $\Op_{eH}$. In particular, through this mixing a contribution to $\kappa_e$ implies a shift in the anomalous magnetic moment of the electron, $\Delta a_e$.

\subsubsection{Indirect constraints from $\Delta a_e$}
To connect to measurements of $\Delta a_e$, we use Eq.~\eqref{eq:eftdipoleRGE} to evolve from the scale of new physics $\Lambda$ to the EW scale $\mu_W$ using the leading-log approximation. At the EW scale, we match onto the coefficient $[L_{e\gamma}]_{ij}$ of the dipole operator $\bar{e}^i_L \sigma_{\mu\nu}e^j_R F^{\mu\nu}$ of the low energy effective field theory (LEFT). The tree level matching gives \cite{Jenkins:2017jig}
\begin{equation}
    L_{e\gamma}(\mu_W)= \frac{v}{\sqrt{2}}(\cos \theta_W\cC_{eB}(\mu_W)-\sin\theta_W\cC_{eW}(\mu_W)).
    \label{eq:LEFT_matching}
 \end{equation}
The contribution to $L_{e\gamma}$ of the two-loop running of Eq.~\eqref{eq:eftdipoleRGE} is then (taking $\mu_W=m_h$):
\begin{equation}
     L^{\text{Running}}_{e\gamma}(m_h) = \frac{3\sqrt{2}}{8}\frac{g^{3}\sin\theta_W \tan^2\theta_W}{(16\pi^2)^2}v\ln\left(\frac{\Lambda}{m_h}\right)\cC_{eH} (\Lambda).\label{eq:Le_running}
\end{equation}

\begingroup
\newcommand{\feylabel}[1]{\Large $#1$}

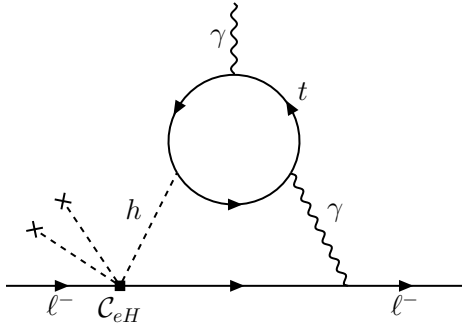
\begin{figure}
    \setlength{\feynhandlinesize}{1pt}
    \centering

        \scalebox{0.75}{\begin{tikzpicture}
            \begin{feynhand}
                \vertex (a) at (0,0);
                \vertex [squaredot, label=below:\feylabel{\mathcal{C}_{eH}}] (b) at (2,0){};
                \vertex (b1) at (1,1.5);
                \vertex (b2) at (0.5,1);
                \vertex (c) at (3,2);
                \vertex (d) at (5,2);
                \vertex (e) at (6,0);
                \vertex (f) at (8,0);
                \vertex (g) at (4,3.73);
                \vertex (h) at (4,5.0);

                \propag[fer] (a) to [edge label' = \feylabel{\ell^-}] (b);
                \propag[sca] (b) to [edge label = \feylabel{h}] (c);
                \propag[fer] (b) to (e);
                \propag[fer] (e) to [edge label' = \feylabel{\ell^-}] (f);
                \propag[bos] (d) to [edge label = \feylabel{\gamma}] (e);
                \propag[fer] (c) to [bend right=60, looseness=1.11] (d);
                \propag[fer] (d) to [bend right=60, looseness=1.11, edge label' = \feylabel{t}] (g);
                \propag[fer] (g) to [bend right=60, looseness=1.11] (c);
                \propag[bos] (g) to [edge label = \feylabel{\gamma}] (h);

                \propag[sca, insertion={[size=3pt]1.0}] (b) to (b1);
                \propag[sca, insertion={[size=3pt]1.0}] (b) to (b2);
            \end{feynhand}
        \end{tikzpicture}}

\caption{An example two-loop Barr-Zee diagram contributing to $\Delta a_\ell$. Crosses on scalar legs indicate Higgs components set to $v$.}
\label{fig:barr_zee_diags}
\end{figure}

\endgroup 
The $\Op_{eH}$ operator also gives a finite matching contribution to $L_{e\gamma}$, arising when the top quark and Higgs, $W$ and $Z$ bosons are integrated out at the EW scale. This contribution proceeds via two-loop Barr-Zee type diagrams \cite{Bjorken:1977vt, Barr:1990vd},
an example of which is shown in Fig.~\ref{fig:barr_zee_diags}. Diagrams involving virtual $W$ and $Z$ boson exchange, in place of the internal photon, are subleading due to their masses and the small vectorial $Z$-lepton coupling \cite{Heinemeyer:2003dq, Heinemeyer:2004yq, Stockinger:2006zn}. To calculate the top loop we map the couplings in the results of \cite{Cheung:2001hz} to the SMEFT framework and obtain 
 \begin{equation}
L_{e\gamma}^{t}(m_h) = \frac{8\sqrt{2}}{3} \frac{g^3\sin^3\theta_W}{(16 \pi^2)^2} v \,f\left(\frac{m_t^2}{m_h^2}\right)
\cC_{eH}(m_h), \label{eq:Le_finite}
\end{equation}
with the loop function $f(z)$ defined as \cite{Stockinger:2006zn}
\begin{equation}
    f(z)=\frac{z(1-2z)}{y} \left[\mathrm{Li}_2 \left(1-\frac{1-y}{2z} \right)- \mathrm{Li}_2 \left(1-\frac{1+y}{2z} \right)\right]+z \left(2+\log z \right),
\end{equation}
where $y=\sqrt{1-4z}$ (note that $f(z)$ is real and analytic even for $z>1/4$).\footnote{Beware that when evaluating this function in \texttt{Mathematica}, the correct result is only found when $1/y$ is coded as \texttt{1/Sqrt[1-4z]}. Instead \texttt{Sqrt[1/(1-4z)]} will give an incorrect overall sign for the first bracket (if $z>1/4$).}
As well as this finite top loop diagram, there is also a divergent Barr-Zee diagram in which the top loop is replaced by a $W$-boson loop. Given that the anomalous dimensions in \eqref{eq:eftdipoleRGE} correspond to the divergent part of this diagram, the finite matching part is scheme-dependent and formally part of the next logarithmic order. We therefore do not include this in the calculation (consistent with the approach in Ref.~\cite{Brod:2022bww}). 

Summing Eqns.\ \eqref{eq:Le_running} and \eqref{eq:Le_finite} we find\footnote{For the sake of compactness, we have omitted the running of $\cC_{eH}$ between $\Lambda$ and the electroweak scale in the matching part of this analytical expression (i.e. we have set $\cC_{eH}(m_h)=\cC_{eH}(\Lambda)$ in the first term when going from \eqref{eq:legamtot1} to \eqref{eq:legamtot2}). This approximation represents at most a 10\% correction to the constraints, but our numerical analysis includes the full running.}
\begin{align}
L_{e\gamma}(m_h) &=L_{e\gamma}^{\text{Matching}}(m_h)+ L_{e\gamma}^{\text{Running}}(m_h)\label{eq:legamtot1} \\
&= \frac{g^3 \sin\theta_W v}{12\sqrt{2}(16 \pi^2)^2}\left(64\sin^2\theta_Wf\left(\frac{m_t^2}{m_h^2}\right)-9\tan^2\theta_W \ln\left(\frac{m_h}{\Lambda}\right)\right)\cC_{eH}(\Lambda).\label{eq:legamtot2}
\end{align}
Finally, neglecting running of $L_{e\gamma}$ below the EW scale,\footnote{The self-renormalisation scaling of this coefficient, within the five-flavour LEFT, is small: $L_{e\gamma}(m_b)\approx0.97L_{e\gamma}(m_h)$ \cite{Fuentes-Martin:2020zaz}.} we can relate this back to $\Delta a_\ell$:
\begin{align}
\Delta a_\ell &= \frac{4m_\ell}{e}\,\mathrm{Re}\,([L_{e\gamma}]_{\ell \ell})\nonumber \\
&= -\frac{g^2m^2_\ell }{3v^2(16 \pi^2)^2}\left(64\sin^2\theta_Wf\left(\frac{m_t^2}{m_h^2}\right)-9\tan^2\theta_W \ln\left(\frac{m_h}{\Lambda}\right)\right)\Delta\kappa_\ell \label{eq:alvskappal},
 \end{align}
where we define $\Delta \kappa_\ell =\kappa_\ell-1$. Current constraints on $\Delta a_e$ are dependent on the experimental value of the fine structure constant $\alpha$. However, there is an existing tension between the two most precise experimental determinations of $\alpha$, obtained using matter-wave interferometry of atomic recoils with either Caesium or Rubidium atoms \cite{Parker:2018vye, Morel:2020dww}. Comparing the resulting SM predictions obtained with the two inputs, to the most recent measurement \cite{Fan:2022eto}, yields \cite{DiLuzio:2024sps}\footnote{The uncertainties on these values, whilst dominated by experimental errors from $\alpha$ determinations and the $a_e$ measurement, receive a non-negligible component from the current $5\sigma$ discrepancy between the two calculations of the contributing 5-loop QED Feynman diagrams without lepton loops \cite{Aoyama:2017uqe, Volkov:2019phy, Volkov:2024yzc}.}
\begin{align}
    \Delta a_e^{\text{Rb}} &= (33.8\pm16.1)\times10^{-14}, \label{eq:aeRb}\\
    \Delta a_e^{\text{Cs}} &= (-102\pm26.4)\times10^{-14} \label{eq:aeCs}.
\end{align}
Clearly, \eqref{eq:aeRb} and \eqref{eq:aeCs} disagree by more than $5\sigma$. In our study, we will take two approaches to dealing with this uncertain situation: (a) taking $\Delta a_e^{\text{Rb}}$ as the current measurement (choosing Rb rather than Cs since it is less discrepant with the SM), and (b) inflating the error bars such that the $2\sigma$ region for $\Delta a_e$ spans the $2\sigma$ region for both determinations. Specifically, for option (b) we define the following $2\sigma$ region:
\begin{equation}
     (\Delta a_e)_{\text{Rb+Cs}}=[-154.8,\,66.0]\times 10^{-14}~ (2\sigma).
\end{equation}
Our qualitative conclusions will generally not depend on which of these approaches we take, but by doing both we can show the impact of the current inconsistency in $\alpha$.

It is expected that by the time of a direct electron Yukawa measurement at \mbox{FCC-ee} there will be roughly an order of magnitude improvement in precision of $\Delta a_e$. With this in mind, we take $\Delta a_e^{\text{future}}< 5\times10^{-14}$ as a rough projection \cite{DiLuzio:2024sps, Erdelyi:2025axy}. Of course, any improvement in precision on $\Delta a_e$ will rely on resolving the discrepancy in $\alpha$. Using Eq.~\eqref{eq:alvskappal}, we calculate the current and future constraints on $\kappa_e$ from $\Delta a_e$ assuming two different new physics scales ($\Lambda=2, 10$ TeV) and list these in Tab.\ \ref{tab:model_indep_constraints}. These are to be compared to the current and future direct $68\%$ limits on $\kappa_e$ in Tab.~\ref{tab:kappa_constraints}. It can be seen that current constraints on $\kappa_e$ from $\Delta a_e$ are very weak, but if a future bound of $\Delta a_e^{\text{future}}< 5\times10^{-14}$ were to be achieved, it could outperform HL-LHC sensitivity.
\begin{table}[h]
    \centering
     \begin{tabular}{|c|c|c|c|}
         \hline
         $\Lambda$ (TeV) & $\kappa_e$ (from $\Delta a_e^{\text{Rb}}$) & $\kappa_e$ (from $\Delta a_e^{\text{Rb+Cs}}$) & $|\kappa_e|$ (projected future $\Delta a_e$)\\
         \hline
         2  & [-890,-320]  & [-890, 2300] & $<46$ \\
         \hline
         10  & [-710, -250] & [-710, 1800] & $<37$ \\
         \hline
     \end{tabular}   
    \caption{68\% confidence intervals for $\kappa_e$ derived from current and future constraints on $\Delta a_e$ under different assumptions for the new physics scale $\Lambda$.}
    \label{tab:model_indep_constraints}
\end{table}

\subsection{Generating $\Op_{eH}$ radiatively}
\label{sec:generating_ceh}
Whilst UV physics could generate $\Op_{eH}$ and therefore modify the electron Yukawa coupling at tree level, the potential precision of future measurements allows us to probe scenarios in which $\Op_{eH}$ arises at loop level. As we found in Sec.~\ref{sec:EFTdipoles}, $\Op_{eH}$ is an element in a set of operators which all mix under RG flow. If any of these operators is generated by integrating out UV physics, the rest of the operators (including $\mathcal \Op_{eH}$) will be populated after RG flow to the IR.
In particular, we identify three semileptonic operators and one four-lepton operator as candidates to radiatively generate $\Op_{eH}$: $\Op^{(1)}_{lequ}, \Op^{(3)}_{lequ}$, $\Op_{ledq}$ and $\Op_{le}$. The dipole operators are the only other possibility, but this possibility is strongly excluded by constraints from $\Delta a_e$. 

The coefficients $\cC^{(1)}_{lequ}$, $\cC_{ledq}$ and $\cC_{le}$ directly enter the one-loop RGE of $\cC_{eH}$~\cite{Jenkins:2013zja,Jenkins:2013wua,Alonso:2013hga}:
\begin{align}
    \frac{d}{d\ln \mu} \cC_{eH} &= \frac{4N_c}{16\pi^2}(Y^\dagger_u Y_u Y_u^\dagger - \lambda Y_u^\dagger)\cC^{(1)}_{lequ} -\frac{4N_c}{16\pi^2}(Y^\dagger_d Y_d Y_d^\dagger-\lambda  Y_d)\cC_{ledq}\nonumber\\
    &+\frac{8}{16\pi^2}(Y^\dagger_e Y_e Y_e^\dagger-\lambda Y_e)\cC_{le},
    \label{eq:ceh_rge}
\end{align}
where flavour indices are suppressed.
The coefficient $\cC^{(3)}_{lequ}$ first mixes into $\cC^{(1)}_{lequ}$ through
\begin{equation}
    \frac{d}{d\ln \mu} \cC^{(1)}_{lequ} = \frac{1}{16\pi^2}(18g^2+30g^{\prime2})\cC^{(3)}_{lequ},
\end{equation}
and subsequently into $\cC_{eH}$. Noting the Yukawa dependence of Eq.~\eqref{eq:ceh_rge}, we study the following coefficient flavour indices:
\begin{equation}
\label{eq:oplist}
    [\cC^{(1)}_{lequ}]_{1133}, [\cC^{(3)}_{lequ}]_{1133}, [\cC^{(1)}_{lequ}]_{1122}, [\cC^{(3)}_{lequ}]_{1122}, [\cC_{ledq}]_{1133}, [\cC_{le}]_{1331},
\end{equation}
where the indices $1,2,3$ correspond to first, second and third generation quarks and leptons respectively.
The flavour charges of these operators, as discussed in Sec.~\ref{sec:EFTdipoles}, indicate we should also expect to generate contributions to the EW dipole operators. Indeed, $\cC_{lequ}^{(3)}$ mixes into $\cC_{eB}$ and $\cC_{eW}$ at one-loop, whilst $\cC^{(1)}_{lequ}$, $\cC_{ledq}$ and $\cC_{le}$ contribute at the two-loop level. So we can again study indirect constraints on $\Delta \kappa_e$ from $\Delta a_e$, assuming that both are generated radiatively from one of the operator coefficients in Eq.~\eqref{eq:oplist}.

Assuming SMEFT coefficients are generated at a new physics scale $\Lambda=2$~TeV, we compute the induced $\cC_{eH}$ through numerically solving the one-loop RGE equations with the \texttt{DsixTools} package~\cite{Celis:2017hod, Fuentes-Martin:2020zaz}.
To calculate contributions to $\Delta a_e$, we perform a resummed one-loop running and one-loop matching analysis, making use of the formulae provided in \cite{Aebischer:2021uvt},\footnote{The contributions of $[\cC^{(1)}_{lequ}]_{1122}$ and $[\cC^{(3)}_{lequ}]_{1122}$ to $\Delta a_e$ depend weakly on unknown non-perturbative charm loop effects, through the parameter denoted $c_T^{(c)}$ in Ref.~\cite{Aebischer:2021uvt}. We set this to zero here for simplicity, but if we were to take it to be $O(1)$, our constraints on these coefficients would change by $O(10\%)$.} supplemented by fixed-order contributions from two-loop anomalous dimensions where available in the literature.
In particular, the calculation of the leading contributions from $[\cC_{ledq}]_{1133}$ and $[\cC_{le}]_{1331}$ to $\Delta a_e$ requires the use of two-loop anomalous dimensions above and below the EW scale. Within the SMEFT, the relevant two-loop RGEs are \cite{Panico:2018hal}:
\begin{align}
\label{eq:eftdipoleRGE4f}
\frac{d}{d\ln \mu}
\left(\begin{array}{c} \cC_{eB}\\ \cC_{eW}\end{array}\right)
&=\frac{Y_dg^3}{(16\pi^2)^2}\frac{N_c}{4}
\left(
\begin{array}{c}
3\tan\theta_W \mathcal{Y}_Q + 4 \tan^3\theta_W (\mathcal{Y}_L +\mathcal{Y}_e)(\mathcal{Y}_Q^2 +\mathcal{Y}_d^2)
\\
\frac{1}{2} + 2 \tan^2\theta_W \mathcal{Y}_Q (\mathcal{Y}_L+\mathcal{Y}_e)
\end{array}
\right)
\cC_{ledq}\nonumber\\
&-\frac{Y_eg^3}{(16\pi^2)^2}\frac{1}{2}
\left(
\begin{array}{c}
3\tan\theta_W \mathcal{Y}_L + 4 \tan^3\theta_W (\mathcal{Y}_L +\mathcal{Y}_e)(\mathcal{Y}_L^2 +\mathcal{Y}_e^2)
\\
\frac{1}{2} + 2 \tan^2\theta_W \mathcal{Y}_L (\mathcal{Y}_L+\mathcal{Y}_e)
\end{array}
\right)
\cC_{le}.
\end{align}
Analogous to the two-loop running calculation in Sec.~\ref{sec:EFTdipoles}, these RGEs are used to calculate the EW dipole coefficients at the electroweak scale, which are then matched to $L_{e\gamma}$ at tree- and one-loop level. $\cC_{ledq}$ and $\cC_{le}$ also have one-to-one tree level matching relations with the four-fermion LEFT coefficients $L^{S,RL}_{ed}$ and $L^{V,LR}_{ee}$ respectively \cite{Jenkins:2017jig}.
For evolution below the EW scale, the two-loop anomalous dimensions for the mixing of these four-fermion LEFT coefficients into $L_{e\gamma}$, have recently appeared in Ref.~\cite{Naterop:2025cwg}:
\begin{equation}
    \frac{d}{d\ln \mu} [L_{e\gamma}]=-\frac{4e^3}{3(16\pi^2)^2}[L^{S,RL}_{ed}][M^\dagger_d]+\frac{50e^3}{9(16\pi^2)^2}[L_{ee}^{V,LR}][M_e^\dagger]
\end{equation}
where $M_d$ and $M_e$ are the diagonal mass matrices for down-type quarks and leptons respectively. We use these to calculate the additional fixed-order contribution to $\Delta a_e$ which is induced from running to the $b$ mass scale.

\begin{figure}
    \centering
    \includegraphics[width=\linewidth]{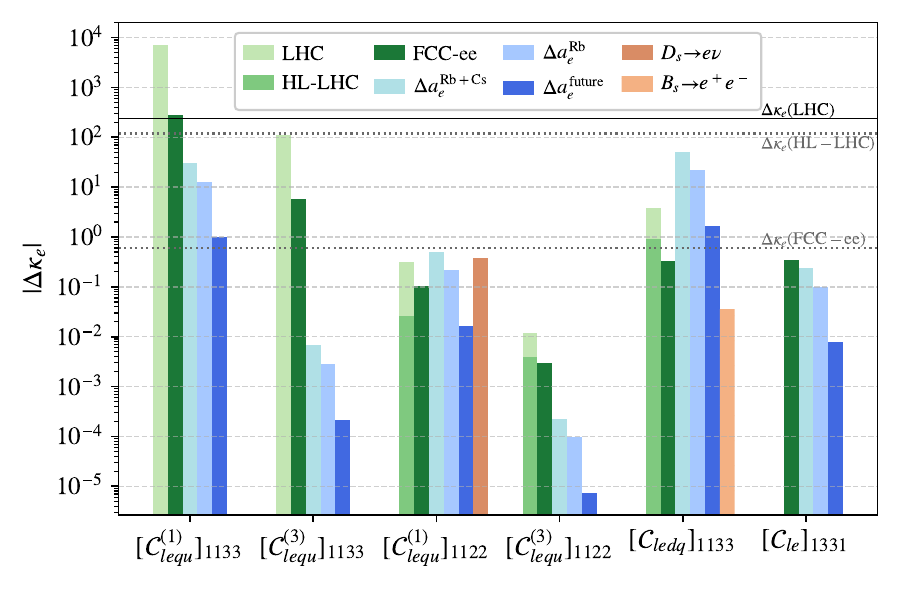}
    \caption{Current and projected upper limits on $|\Delta \kappa_e|$(defined as $|\kappa_e-1|$) at 95\% CL, assuming coefficients are generated at 2~TeV and only one SMEFT coefficient is non-zero at a time. Current LHC bounds on $[\cC_{lequ}^{(1)}]_{1133}$ and $[\cC_{lequ}^{(3)}]_{1133}$ are taken from \cite{CMS:2023xyc}, while those on $[\cC_{lequ}^{(1)}]_{1122}$, $[\cC_{lequ}^{(3)}]_{1122}$ and $[\cC_{ledq}]_{1133}$ are from \cite{Allwicher:2022gkm}. All HL-LHC and FCC-ee projections, shown by green bars, are taken from \cite{Greljo:2024ytg}. Constraints from the decays $D_s^+\to e^+\nu$ and $B_s \to e^+e^-$ follow from App.~\ref{sec:mesondecays}, where the hatching on the latter bar indicates the assumption that the coefficient is in the up mass basis. Future $\Delta a_e$ sensitivity is assumed to reach $\Delta a_e^{\text{future}} < 5 \times 10^{-14}$.}
    \label{fig:generating_ceh_from_running} 
\end{figure}

Putting this all together, assuming only one non-zero SMEFT coefficient at a time, Fig.~\ref{fig:generating_ceh_from_running} shows their contributions to $\Delta \kappa_e$, as constrained by current and projected bounds on $\Delta a_e$, as well as by complementary searches at existing and proposed colliders. 
We observe a strong degree of complementarity between collider and $\Delta a_e$ constraints on the relevant SMEFT coefficients. Under current collider limits the coefficients $[\mathcal{C}^{(1)}_{lequ}]_{1133}$ and  $[\mathcal{C}^{(3)}_{lequ}]_{1133}$ can accommodate $\kappa_e$ enhancements at the level of HL-LHC sensitivity; however, present $\Delta a_e$ measurements already exclude this possibility. If the HL-LHC were to observe a deviation $\Delta\kappa_e \sim 120$, current $\Delta a_e$ bounds would therefore point to UV physics matching directly to $[\mathcal{C}_{eH}]_{11}$.

For some of the coefficients in the list \eqref{eq:oplist}, specifically ones which do not involve a top quark, there are additional constraints from meson decays to electrons, to which they contribute with a large chiral enhancement. This applies to the coefficient $[C_{ledq}]_{1133}$ which can mediate the $B_s\to e^+e^-$ decay (if the quark doublet flavour index is in the up mass basis), and the coefficient $[C_{lequ}^{(1)}]_{1122}$ which can mediate the $D_s^+\to e^+\nu$ decay. Neither of these decay modes have been measured, and the experimental upper limits on them are orders of magnitude above the SM prediction~\cite{LHCb:2020pcv,Belle:2013isi}, but the chiral enhancement of the NP contribution from these coefficients is so large that meaningful constraints can nevertheless be set from these decays. The details of the calculation of these constraints are given in App.~\ref{sec:mesondecays}. In Fig.~\ref{fig:generating_ceh_from_running}, we show the corresponding upper limits on $\Delta \kappa_e$ in orange, assuming that a deviation in $\kappa_e$ arises radiatively from the given coefficient. The crosshatching on the bar for $[C_{ledq}]_{1133}$ is to indicate that this bound depends on the flavour alignment of the operator indices; if the quark doublet flavour index is defined in the up type quark mass basis, then the constraint applies, but if it is defined in the down type quark mass basis, then there is no down-type flavour change and the constraint is entirely evaded. By contrast, the bound on $[C_{lequ}^{(1)}]_{1122}$ involves a charged current, and is almost independent of whether the indices are defined in the up or down mass basis (the difference is a factor of $V_{cs}\approx 0.97$).

Two coefficients remain viable candidates for radiatively inducing $\Delta\kappa_e$ above the FCC-ee Higgs-pole sensitivity without violating current bounds: $[\cC^{(1)}_{lequ}]_{1133}$ and $[\cC_{ledq}]_{1133}$ (if aligned in the down-type quark mass basis). The parameter space for the latter could be fully tested by measurements of hadronic ratios above the $Z$-pole \cite{Greljo:2024ytg}. This leaves $[\cC^{(1)}_{lequ}]_{1133}$, where the relevant parameter space for generating $\Delta \kappa_e \gtrsim 0.6$ can be largely probed by a future measurement of $\Delta a_e \lesssim 5 \times 10^{-14}$. We conclude that at the time of an FCC-ee Higgs-pole run, a $\kappa_e$ enhancement at this level, in the absence of anomalies in other processes, would again favour UV completions generating $[\cC_{eH}]_{11}$ at leading order.

We note the simplification of studying only one SMEFT operator non-zero at any time could obscure scenarios in which large $\kappa_e$ enhancements are allowed through the interplay of multiple SMEFT coefficients. Furthermore, in complete models one expects finite matching contributions at $\Lambda$ to arise in similar places to those arising from RG evolution, thus an overly strict numerical interpretation of the results is to be discouraged. We study motivated multi-coefficient scenarios in the form of single field SM extensions in Sec.~\ref{sec:models}. 

\subsection{Constraints on the flavour structure of $\mathcal O_{eH}$}
Although our goal is to specifically enhance the electron Yukawa by imposing a non-zero $[\cC_{eH}]_{11}$, we are still required to make some assumption about the rest of the flavour entries of the coefficient, which are not restricted by any symmetry \textit{a priori}. Obviously, the least constrained flavour structure would be one in which all of the other entries of $\cC_{eH}$ are zero, which we denote the `electrophilic' structure.  However, such a scenario appears contrived from a UV perspective and so we ought to consider a range of possible UV scenarios.  

In this section we review a few examples of the various flavour structures that can arise. We start with generic flavour constraints, without making any assumptions about $\cC_{eH}$, and then examine more specific structures and in particular ones which avoid flavour-changing neutral currents (FCNCs) and correlate the diagonal components of $\cC_{eH}$.

To review, both the lepton mass matrix and the CP-even Higgs Yukawas receive contributions from $\mathcal O_{eH}$. We may diagonalise the lepton mass matrix with unitary flavour rotations $U_l$ and $U_E$,
\begin{align}
    l \to U_l l ~~~,~~~
    e \to U_e e \,,
\end{align}
resulting in Higgs couplings in the mass basis
\begin{align}
    \mathcal{L}_h &\supset - [y_e^{\rm eff}]_{ij} h \bar \ell_L^i \ell_R^j, \\
    y_e^{\rm eff} &= \frac{M_{ll}}{v} - \frac{v^2}{\sqrt{2}} U_l^\dagger \cC_{eH} U_e,
\end{align}
where we have used $\ell$ to indicate the charged lepton after electroweak symmetry breaking, and $M_{ll}$ is the diagonalised charged lepton mass matrix.  Note that unless $y_e$ and $\cC_{eH}$ were already aligned in flavour space there is no reason to expect the combination $U_\ell^\dagger \cC_{eH} U_e$ to be a diagonal matrix. 

\subsubsection{Experimental probes of off-diagonal elements of $\mathcal O_{eH}$}

There are two experimental channels that constrain the lepton flavour violating (LFV) Higgs couplings, and they offer complementary bounds. 
For one, there are the LHC searches for LFV Higgs decays, where a Higgs from a traditional production channel decays to a pair of leptons of differing flavour. This search channel offers a direct probe of the lepton Yukawas at tree-level, and the subsequent runs at the LHC will steadily improve this bound. These bounds are most important in probing the LFV $\tau$ couplings. 

\begin{table}[h]
    \centering
    \begin{tabular}{|c|c|c|c|}
    \hline
         & Current (PDG) & HL-LHC\,\cite{ATL-PHYS-PUB-2022-054} & FCC\,\cite{Qin:2017aju}\\ \hline
         BR($h\to \mu e$) & $4.4\times 10^{-5}$ & - & $1.2\times 10^{-5}$  \\
         BR($h\to \tau e$) & $2.0\times 10^{-3}$ & $2.4\times 10^{-4}$ & $1.6\times 10^{-4}$ \\
         BR($h\to \tau \mu$) & $1.5\times 10^{-3}$ & $2.4\times 10^{-4}$  & $1.4\times 10^{-4}$  \\
         \hline
    \end{tabular}
    \caption{Current and future 95\% CL expected limits on charged LFV Higgs decays.}
    \label{tab:hlfv}
\end{table}
\noindent
The branching ratio of $h\to \ell_i\ell_j$ in the SMEFT, with $i\neq j$, is given by
\begin{align}
    \mathcal{B}(h\to\ell_i\ell_j) = \frac{1}{\Gamma_h} \frac{m_h}{16\pi} v^4 \left(|[\cC_{eH}]_{ij}|^2 + |[\cC_{eH}]_{ji}|^2\right) \,,
\end{align}
where we have neglected lepton masses in the kinematics.
The current and future experimental constraints are summarised in Table \ref{tab:hlfv}.

Alternatively, radiative LFV decays of leptons receive loop contributions involving the Higgs, thereby accessing the coefficients $\cC_{eH}$. Because $\tau$s are somewhat difficult to produce and observe, the $\tau$ LFV decay searches are subdominant to the relevant LHC constraints. In contrast, the naturally long lifetime of the muon makes it an extremely sensitive probe of EW-scale physics, and in turn provides a strong constraint on flavour violation in the $\mu e$ sector. The sensitivity of these searches will improve considerably by the time the FCC becomes operational. We therefore review these constraints in brief, and then use them to evaluate a simple flavour scenario in the following section. The outlook of the future of muonic decay experiments, as well as their projected constraining power, is summarised in Table \ref{tab:mu_to_e}. 

\begin{table}[t]
    \renewcommand{\arraystretch}{1.1}
    \centering
    \begin{tabular}{|c | c |c| c |c |c|}
    \hline
        &&& \multicolumn{2}{c|}{Limit on $\frac{v^2}{y_e^{\text{SM}}}[\cC_{eH}(m_h)]_{\langle e\mu \rangle}$} \\
        Experiment & Limit & Date & \multicolumn{1}{c}{ $\Lambda = 2$\,TeV} & $\Lambda = 10\,$TeV \\ \hline
        Sundrum-II \cite{SINDRUMII:2006dvw} & BR($\mu\,\text{Au} \to e\,\text{Au}) < 7 \times 10^{-13} $ & Current & 6.6 & 6.0 \\
        MEG-II \cite{MEGII:2025gzr} & BR($\mu \to e \gamma) < 1.5 \times 10^{-13}$ & Current & 0.39 & 0.33 \\
        MEG-II  \cite{MEGII:2025gzr} & BR($\mu \to e \gamma) < 6 \times 10^{-14}$ & Late 2020s & 0.25 & 0.21 \\
        Mu2e \cite{Mu2e:2022ggl} & BR($\mu\,\text{Al} \to e\,\text{Al}) < 6 \times 10^{-16}$ & Late 2020s & 0.22 & 0.20\\
        Mu3e \cite{Berger:2014vba} & BR($\mu \to 3 e) \lesssim \mathcal O(10^{-16})$ & $\sim$2030 & $\sim$0.10 & $\sim$0.08 \\
        COMET \cite{Moritsu:2022lem}  & BR($\mu\,\text{Al} \to e\,\text{Al}) \lesssim \mathcal O(10^{-17})$ & Late 2030s & $\sim$0.03 & $\sim$0.03\\
        \hline
    \end{tabular}
    \caption{Summary of limits on $\cC_{eH}$ originating from searches for rare muon decays, as well as projections for future searches, all reported at 90\%~CL. For compactness, we denote the flavour average $C_{\langle e \mu \rangle} \equiv \sqrt{|C_{e\mu}|^2 + |C_{\mu e}|^2}$.} 
    \label{tab:mu_to_e}
\end{table}

As discussed in Section~\ref{sec:EFTdipoles}, the photonic dipole coefficient $L_{e\gamma}$ is induced at two loops through running and matching contributions from $\cC_{eH}$. The LFV decay probes are also somewhat sensitive to $\kappa_t$ and $\kappa_W$, but motivated by the relevant LHC constraints we assume that they are effectively unity. The rate for the LFV decay $\ell_i \to \ell_j \gamma$ is~\cite{Harnik:2012pb}
\begin{align}
    \Gamma(\ell_i \to \ell_j \gamma) = \frac{m_{i}^3}{4\pi} \left( \left| [L_{e\gamma}(m_h)]_{ij} \right|^2 + \left| [L_{e\gamma}(m_h)]_{ji} \right|^2\right).
\end{align}
Current searches for the rare decay $\mu\to e\gamma$ offer the strongest bound on the $\mu e$ and $e\mu$ entries of $L_{e \gamma}$, set by the MEG-II experiment at BR($\mu \to e\gamma) < 1.5 \times 10^{-13}$ \cite{MEGII:2025gzr}. In future years, other search channels will overtake this bound, so we briefly describe these decay modes.

Another important search channel is the  $\ell_i \to 3 \ell_j$ decay mode, which is mediated by a virtual photon or Higgs. The Higgs-mediated diagram is subdominant in comparison to the photon-mediated diagram due to the small electron Yukawa and it can be omitted at leading order. The photon-mediated diagram contains the same effective $L_{e\gamma}$ coupling, and the rate is approximately \cite{Harnik:2012pb}
\begin{align}
    \Gamma(\ell_i \to 3 \ell_j) \simeq \frac{ \alpha m_i^3}{12\pi^2} \left|\log\frac{m_j^2}{m_i^2} - \frac{11}{4}\right| \times \left( \Big| [L_{e\gamma}(m_h)]_{\mu e} \Big|^2 + \Big| [L_{e\gamma}(m_h)]_{e\mu} \Big|^2\right).
\end{align}
Note that this rate is suppressed over $\mu \to e \gamma$ by an additional factor of $\alpha$. Nevertheless, the upcoming Mu3e experiment is expected to probe branching ratios for $\mu \to 3e$ down to $\lesssim 10^{-16}$~\cite{Berger:2014vba}, improving the current bound on the $\mu e$ couplings of $\cC_{eH}$ by a factor of a few.

Lastly, $\mu \to e$ conversion is expected to compete with $\mu \to 3e$ and potentially result in the strongest bound on $\mu e$ LFV, improving current constraints by up to two orders of magnitude. Both Higgs- and photon-mediated diagrams contribute to $\mu \to e$ conversion but, in contrast to the $\mu \to 3e$ case, we find the Higgs-mediated diagram is numerically larger. This is because the effective Higgs coupling to quarks in the nucleus is much larger than the electron Yukawa, which relatively suppresses $\mu \to 3e$. For further details on this calculation see App.~\ref{app:mu2eConversion}.

To summarise, we compile the current and projected strongest constraints on $\cC_{eH}$ (from both Higgs decays in Tab.~\ref{tab:hlfv} and Tab.~\ref{tab:kappa_constraints}, and $\mu\to e$ transitions in Tab.~\ref{tab:mu_to_e}) into the following matrices (at 95\% CL), with symmetric off-diagonal entries:
\begin{align}
    |\cC_{eH} (m_h)| \times {\rm TeV}^2 \lesssim
    \begin{cases}
        \begin{pmatrix}
        1.2 \times 10^{-2}& 1.1 \times 10^{-5} & 3.0\times 10^{-2} \\
        & 6.1 \times 10^{-3} & 2.6\times 10^{-2} \\
        & &  2.3 \times 10^{-2}
    \end{pmatrix} & \text{(current)}
    \\
        \begin{pmatrix}
        7.4 \times 10^{-5}& 9.2 \times 10^{-7} & 8.5\times 10^{-3} \\
        & 7.7 \times 10^{-4} & 7.9\times 10^{-3} \\
        & &  2.2 \times 10^{-3}
    \end{pmatrix} &\text{(projected)}
    \end{cases}
\end{align}
where flavour indices run over $e$, $\mu$, $\tau$ in that order from left to right and top to bottom. In the $e\mu$ entry, we have calculated the indirect limits assuming a new physics scale $\Lambda = 10$\,TeV. All other entries are obtained from direct bounds on $h\to f f^\prime$ or $e^+e^-\to h$.

\subsubsection{Examples of possible flavour structures}
\paragraph{Anarchic Flavour}
The first and most straightforward UV possibility is one of full flavour anarchy, where all the entries of $\cC_{eH}$ are of the same order. 
We assume that the anarchic structure of $\cC_{eH}$ survives lepton mass diagonalisation, which is natural if the lepton mass matrix is mainly set by the renormalisable Higgs Yukawas. In this scenario, the strongest constraint on any entry of $\cC_{eH}$ is set by the current bound on the branching ratio of $\mu \to e\gamma$ from Meg-II (Table \ref{tab:mu_to_e}), leading to a rough 95\% CL indirect bound on $\kappa_e$ of
\begin{align}
    |\Delta \kappa_e| = \frac{v^3}{\sqrt 2 m_e} \Big|[\cC_{eH}]_{ee} \Big| \sim  \frac{ v^3}{\sqrt 2m_e} \Big|[\cC_{eH}]_{e\mu} \Big| \lesssim 0.3 \qquad (\text{Anarchic flavour}).
\end{align}
An $\mathcal O(1)$ shift in $\kappa_e$ is therefore allowed under an anarchic scenario without violating current flavour constraints or requiring excessive tuning.
However, future improvements in muonic decay searches may improve this bound significantly prior to the start of the FCC. In particular, if COMET achieves its long-term sensitivity goals, the corresponding bound on $\Delta \kappa_e$ could be tightened to $\lesssim 0.02$, effectively excluding the possibility of observing it at the FCC-ee Higgs pole run under the anarchic flavour assumption. 

\paragraph{Aligned flavour}

In the absence of neutrino masses, the SM possesses an exact vectorial $U(1)_e\times U(1)_\mu\times U(1)_\tau$ lepton flavour symmetry. It also possesses an approximate larger vectorial $U(3)_D$
symmetry which is broken only by the differences between the lepton Yukawa couplings. We can imagine that BSM physics also obeys these symmetries, giving flavour-aligned Wilson coefficients in both cases, and universal (and diagonal) Wilson coefficients in the latter. In the presence of neutrino masses, the misalignment between the neutrino mass basis and the lepton mass basis is physical, and it is reasonable to ask why effects of heavy BSM should be aligned with the lepton mass basis. Mechanisms such as a leptonic version of Spontaneous Flavour Violation (SFV) \cite{Egana-Ugrinovic:2018znw} may exist to justify this. Such an alignment in the lepton sector is radiatively stable (up to tiny neutrino mass effects), and ensures that charged lepton flavour violating operators are negligible, being suppressed by powers of $m_\nu/v$.

Among the possible flavour alignments we wish to highlight a few interesting options.

\begin{table}
\centering
\begin{tabular}{|l | c | c | c | c|}
\hline
Coupling modifier & Current & HL-LHC & FCC-ee & FCC-hh \\
\hline 
        $|\kappa_e|$ & $\lesssim 120$ \cite{CMS:2022urr} & $\lesssim 60$ \cite{Cepeda:2019klc} & $\lesssim 1.3$ \cite{FCC:2025lpp} & --- \\
        $\kappa_\mu$ & $1.07^{+0.25}_{-0.31}$ \cite{ATLAS:2022vkf} & $\pm 0.028$ \cite{deBlas:2944678}  & $\pm 0.023$ \cite{deBlas:2944678} & $\pm 0.004$ \cite{deBlas:2944678} \\
        $\kappa_\tau$ & $0.93\pm 0.07$ \cite{ATLAS:2022vkf} & $\pm 0.016$ \cite{FCC:2025lpp} & $\pm 0.0046$ \cite{FCC:2025lpp}  &  $\pm 0.004$ \cite{FCC:2025lpp}\\
       \hline
\end{tabular}
\caption{Current constraints and future sensitivity (all at 68\%~CL) on Higgs-charged lepton coupling modifiers.  The FCC-ee projection assumes a Higgs-pole run and there is no projection for FCC-hh available.  Where necessary 95\%~CL projections have been halved to estimate the 68\%~CL sensitivity.
}
\label{tab:kappa_constraints}
\end{table}

\begin{itemize}
\item \textbf{Electrophilic}
The simplest possibility is that the UV serendipitously features a form of flavour alignment and `diagonality' such that in the basis in which the charged lepton Yukawa couplings have been diagonalised the $\mathcal{O}_{eH}$ operator generated in the UV lies entirely in the electron flavour direction. In the absence of neutrino masses, this can be achieved by charging the BSM physics under $U(1)_e$ only.

In this case, within the EFT the only additional constraint on $C_{eH}$ other than $\kappa_e$ itself is the anomalous magnetic moment of the electron, as discussed in Sec.~\ref{sec:EFTdipoles}.

\item \textbf{Universal}
This is the case for which $U_l^\dagger \cC_{eH} U_e \propto I_3$, i.e. the UV breaks the flavour symmetry as $\text{SU}(3)_L \times \text{SU}(3)_E \to \text{SU}(3)_D$, i.e.\ to the diagonal subgroup, predicting
\begin{equation}
    \Delta \kappa_e = \Delta \kappa_\mu \frac{m_\mu}{m_e} = \Delta \kappa_\tau \frac{m_\tau}{m_e} ~~.
\end{equation}
Clearly the electron Yukawa modifications can, in this instance, be significant as compared to the heavier leptons.  However note in Tab.~\ref{tab:kappa_constraints} that at HL-LHC (FCC-hh) the projected future muon Yukawa constraints are at the $4\%$ ($0.4\%$) level.  The corresponding indirect limit on the electron Yukawa modification, under this ansatz, would then be $\Delta \kappa_e \lesssim 8~(0.8)$ at 68\%~CL.

\item \textbf{Minimal flavour violation (MFV)}
In this case $U_l^\dagger \cC_{eH} U_e \propto \hat M$, such that 
\begin{equation}
    \Delta \kappa_e = \Delta \kappa_\mu = \Delta \kappa_\tau ~~.
\end{equation}
With $\kappa_\tau$ presently constrained as in Tab.~\ref{tab:kappa_constraints}, any possible deviations in the electron Yukawa would be correspondingly small.  If the UV is MFV-like it is unlikely that electron Yukawa measurements would ever be competitive with the respective $\tau$ constraints.
\end{itemize}
To summarise, under the well-motivated Anarchic, MFV and Universal UV flavour assumptions we find that $\mu\to e \gamma$, HL-LHC and FCC-hh, would provide superior or competitive indirect bounds on the electron Yukawa coupling, respectively, as compared to an FCC-ee Higgs pole run.  A more specific electrophilic scenario must be envisaged in order for a significantly modified electron Yukawa coupling to evade indirect constraints achievable elsewhere.

\section{Selected extensions which enhance the electron Yukawa}
\label{sec:models}

We now move from the EFT description to discuss simple extensions of the SM that can enhance the electron Yukawa. Within explicit models, more operators are generated than just $\Op_{eH}$, meaning that additional tests and constraints can be envisaged, dependent on the individual model. However, we can again use the fact that the new physics must provide an additional source of electron chiral symmetry breaking to make some general statements. 

In particular, if the contribution to the electron Yukawa from some model is dependent on a particular product of BSM couplings, then that same product of couplings can only appear in other observables that also break electron chiral symmetry. This can be understood through spurionic or flavour symmetry arguments similar to those in Sec.~\ref{sec:EFTdipoles}. Observables in which a vector current of leptons appears in the amplitude will instead be dependent on a different product of couplings. So, even within explicit models in which other effects are generated, the electron magnetic dipole moment is the only observable we can expect to be entirely correlated with the electron Yukawa. 

For this reason, we begin this section with a discussion of the expected loop suppression of the dipole operators relative to the Yukawa operators in BSM models. We then explore simple single-particle extensions in more detail; a second Higgs doublet $\varphi$, and scalar leptoquarks.

\subsection{Model-dependent connections between Yukawas and dipoles}
\label{sec:general_connection}
\begin{figure}
    \centering
    \begin{tikzpicture}
   \begin{feynhand}
   \setlength{\feynhandblobsize}{10mm}
   \vertex[label=$L$] (a) at (-2,-0.5);      
\vertex [grayblob] (b) at (0,0) {};
  \vertex[label=$e$] (c1)  at (2,-0.5);   
   \vertex[label=$H$] (c2) at (-1.5,1);
   \vertex[label=$H^\dagger$] (c3) at (0,1.5);
    \vertex[label=$H$] (c4) at (1.5,1);
   \propag [fer] (a) to (b);
   \propag [fer] (b) to (c1);
   \propag [sca] (b) to (c2);
    \propag [sca] (b) to (c3);
     \propag [sca] (b) to (c4);
   \end{feynhand}
\end{tikzpicture}~~~~~~~
    \begin{tikzpicture}
   \begin{feynhand}
   \setlength{\feynhandblobsize}{10mm}
   \vertex[label=$L$] (a) at (-2,-0.5);      
\vertex [grayblob] (b) at (0,0) {};
   \vertex[label=$e$]  (c1) at (2,-0.5);   
   \vertex[label=$H$] (c2) at (-1.5,1);
    \vertex (d) at (0.8,0.7);
    \vertex (e)  at (1.0,1.1);
    \vertex[label=$B/W$] (f) at (1.8,1.7);
   \propag [fer] (a) to (b);
   \propag [fer] (b) to (c1);
 \propag [sca] (b) to (c2);
   \propag [sca] (b) to [out=0, in=300] (d);
   \propag [sca] (d) to [out=140, in=80] (b);
   \propag [bos] (f) to (e);
   \end{feynhand}
\end{tikzpicture}
    \caption{Schematic diagrams showing how the EW dipole operators $O_{eW}$ and $O_{eB}$ are generically generated at one loop order higher than the $O_{eH}$ operator in UV models. The grey blob represents a diagram of arbitrary loop order involving exchange of heavy UV states. The left hand diagram matches to the dimension-six $O_{eH}$ SMEFT operator, while the right hand diagram matches to the dimension-six dipole operators. The gauge boson line connects to any charged particle in the diagram, including the $H$ loop and any charged particles within the grey blob.}
    \label{fig:loopg-2}
\end{figure}
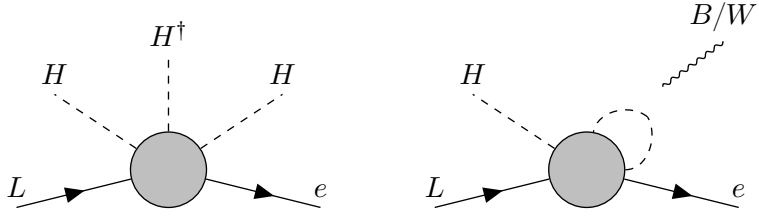

As noted in Sec.~\ref{sec:EFTdipoles}, there is a 2-loop connection within the EFT between the Yukawa-like operators and the dipole operators, which allows weak but model-independent constraints to be put on deviations in $\kappa_e$ from measurements of $\Delta a_e$. However, in almost every UV completion, this connection arises at lower loop order. Schematically, this can be understood by the diagrams in Fig.~\ref{fig:loopg-2}; for any UV completion it is generally possible to close a Higgs loop and radiate a gauge boson to obtain the electroweak dipole operators at one loop order higher than the $\cC_{eH}$ operator.

\begin{table}[]
    \centering
    \begin{tabular}{|l|c | l|  c|c | c|}
    \hline
    State & Spin & SM charges  & $\cC_{eH}$ & $\cC_{e\gamma}$ & Coupling dependence\\
    \hline
    \textcolor{blue}{$\mathcal{S}$} & 0 & $(1,1,0)$ & tree & 1 loop  & $\kappa_\mathcal{S}(\tilde y^e_\mathcal{S})^*_{11}/f$\\
    $\varphi$  & 0 & $(1,2,\frac{1}{2})$ & tree & 2 loop & $\lambda_\varphi (y^e_\varphi)^*_{11} $\\
    \textcolor{blue}{$\Xi$} & 0 & $(1,3,0)$ & tree & 1 loop  & $\kappa_\Xi(\tilde y^e_\Xi)^*_{11}/f$\\
    \textcolor{blue}{$\Xi_1$} & 0 & $(1,3,1)$ & tree & 1 loop  & $\kappa_{\Xi_1}(\tilde y^e_{\Xi_1})^*_{11}/f$ \\
    \textcolor{blue}{$E$} & $\frac{1}{2}$ & $(1,1,-1)$ & tree & 1 loop  & $(\tilde \lambda_E^e)_1(\lambda_E)^*_1/f$\\
    \textcolor{blue}{$\Delta_1$} & $\frac{1}{2}$ & $(1,2,-\frac{1}{2})$ & tree & 1 loop  & $(\tilde \lambda_{\Delta_1}^l)^*_1(\lambda_{\Delta_1})_1/f$\\
    \textcolor{blue}{$\Delta_3$} & $\frac{1}{2}$ & $(1,2,-\frac{3}{2})$ & tree & 1 loop & $(\tilde \lambda_{\Delta_3}^l)^*_1(\lambda_{\Delta_3})_1/f$ \\
    \textcolor{blue}{$\Sigma$} & $\frac{1}{2}$ & $(1,3,0)$ & tree & 1 loop  & $(\tilde \lambda_\Sigma^e)_1(\lambda_\Sigma)^*_1/f$\\
    \textcolor{blue}{$\Sigma_1$} & $\frac{1}{2}$ & $(1,3,-1)$ & tree & 1 loop & $(\tilde \lambda_{\Sigma_1}^e)_1(\lambda_{\Sigma_1})^*_1/f$ \\
    \textcolor{blue}{$\mathcal{L}_1$} & $1$ & $(1,2,\frac{1}{2})$ & tree$^\ast$ & tree$^\ast$ & $\left\{ (\tilde{g}^{Del}_{\mathcal{L}_1})^*_{11} ,~ (\tilde{g}^{eDl}_{\mathcal{L}_1})^*_{11} \right\}\gamma_{\mathcal{L}_1}/f$\\    
    \hline 
    $\varphi$  & 0 & $(1,2,\frac{1}{2})$ & 1 loop & 2 loop & $(y^e_\varphi)^*_{11}(y^u_\varphi)_{33}$ \\
    $\omega_1$ & 0 & $(3,1,-\frac{1}{3})$ & 1 loop & 1 loop &$(y_{\omega_1}^{eu})_{13}(y_{\omega_1}^{ql})^*_{31}$\\
    $\Pi_7$ & 0 & $(3,2,\frac{7}{6})$ & 1 loop & 1 loop & $(y_{\Pi_7}^{eq})^*_{13}(y_{\Pi_7}^{lu})_{13}$\\
    $\mathcal{U}_2 $ & 1 & $(3,1,\frac{2}{3})$ & 1 loop & 1 loop & $(g_{\mathcal{U}_2}^{ed})^*_{13}(g_{\mathcal{U}_2}^{lq})_{13}$\\
    $\mathcal{Q}_5 $ & 1 & $(3,2,-\frac{5}{6})$ & 1 loop & 1 loop & $(g_{\mathcal{Q}_5}^{eq})^*_{13}(g_{\mathcal{Q}_5}^{dl})_{31}$\\
    \hline
    \end{tabular}
    \caption{States which match at tree or one loop level onto $\cC_{eH}$, with coefficients unsuppressed by $y_e$ (or by any other SM Yukawa couplings smaller than $y_b$). In all cases, $\cC_{e\gamma}$ is generated via the same product of couplings as $\cC_{eH}$, with this coupling dependence given in the last column (in the notation of \cite{deBlas:2017xtg}). States labelled in \textcolor{blue}{blue} match to $\cC_{eH}$ (and $\cC_{e\gamma}$) via diagrams involving a non-renormalisable interaction, meaning that the UV completion of these operators would require an additional BSM state. The asterisks on `tree$^{*}$' for the $\mathcal{L}_1$ state indicate that this state matches at tree level onto these operators, but that the couplings involved cannot arise within a weakly coupled UV completion (for example it is known that $\mathcal{C}_{e\gamma}$ cannot be generated at tree level within a weakly coupled UV completion~\cite{Craig:2019wmo,Grojean:2024tcw}).
    }
    \label{tab:UVcompletions}
\end{table}

For individual states which generate $\cC_{eH}$ at tree or 1 loop level, this can be seen explicitly model-by-model, as summarised in Tab.~\ref{tab:UVcompletions}. Here we have defined $\cC_{e\gamma}$ as the coefficient of the SMEFT photonic dipole operator $(\bar{\ell}\sigma_{\mu\nu}e)H F^{\mu\nu}$, which is given in terms of $\cC_{eB}$ and $\cC_{eW}$ as $\cC_{e\gamma}=\cos \theta_W \cC_{eB}-\sin\theta_W \cC_{eW}$. In all but one case listed in this table, $\cC_{e\gamma}$ is generated at one loop order higher than $\cC_{eH}$, or they are both generated at the same loop order. 
The exception is the $SU(2)_L$ doublet scalar $\varphi$, 
which has a 2 loop suppression of $\cC_{e\gamma}$ relative to $\cC_{eH}$ (in the case where it has no coupling to tops). This can be understood by the fact that, since all the Higgs legs in the tree level diagram generating $\cC_{eH}$ come from the same vertex, closing a Higgs loop in this case gives a zero amplitude by the Ward identity:
\begin{equation}
\begin{tikzpicture}[baseline=4ex]
    \begin{feynhand}
        \vertex[label=$\varphi$] (a) at (-2.0,0.0);
        \vertex[dot] (x) at (0.0,0.0) {};
        \vertex[label=$H$] (b) at (2.0,0.0);
        \vertex (e) at (0.0,1.5);
        \vertex (f) at (0.7,2.2);
        \propag[sca] (a) to (x);
        \propag[sca] (x) to (b);
        \propag[sca,mom={}] (x) to [out=0, in=0,looseness=1.5,edge label'=$k$] (e);
        \propag[sca,mom={}] (e) to [out=180, in=180,looseness=1.5,edge label'=$k-q$] (x);
        \propag[bos,mom={$q$}] (e) to (f);
    \end{feynhand}
\end{tikzpicture}~~ \propto~ q \cdot \epsilon^*(q) =0.
\end{equation}
The only particles which have a renormalisable and gauge invariant interaction with three $H$ bosons are the $SU(2)_L$ doublet scalar $\varphi$ and the $SU(2)_L$ quadruplet scalars $\Theta_1$ and $\Theta_3$. Therefore only models including these states can allow for $\Op_{e\gamma}$ to be generated at 2 loop orders higher than $\Op_{eH}$.
Every other UV completion will match to $\Op_{eH}$ with a diagram in which at least one $H$ leg will originate from a different vertex to the $H^\dagger$ leg, thereby avoiding the kinematic situation which gives zero in $\Op_{e\gamma}$ upon closing a Higgs loop. 

Schematic indirect bounds on $\Delta \kappa_e$ from current and future measurements of $\Delta a_e$ are shown in Tab.~\ref{tab:schematicboundsloops}, for the three possible options for the relative loop suppression of $\cC_{e\gamma}$ compared to $\cC_{eH}$. These are calculated under the simple assumption that 
\begin{equation}
C_{e\gamma}=\frac{e}{16\pi^2}\left(\frac{g^2}{16\pi^2}\right)^{n-1} \cC_{eH} ~~,
\end{equation}
where $n=0,1,2$ is the relative loop suppression, and $g$ is some SM or BSM coupling which we take to be $O(1)$. In many realistic cases there would also be a logarithmic enhancement, which may render these estimates overly conservative. 

\begin{table}
\centering
\begin{tabular}{|c|c|c|c|}
         \hline
         Relative loop suppression & $\Delta\kappa_e$ ($(\Delta a^{\text{Rb}}_e)$) & $\Delta\kappa_e$ ($(\Delta a^{\text{Rb+Cs}}_e)$) & $|\Delta\kappa_e|$ ($\Delta a^{\text{future}}_e$)\\
         \hline
         0  & [-0.004, -0.20] & [-0.20, 0.40] & $<0.01$ \\
         \hline
         1  & [-0.6, -20] & [-20, 60] & $<2$ \\
         \hline
         2 & [-90, -4000] & [-4000, 9000] & $<300$ \\
         \hline
     \end{tabular}   
     \caption{\label{tab:schematicboundsloops}Estimated constraints at 95\% confidence on $\Delta \kappa_e$ from current and future $\Delta a_e$ measurements, considering different scenarios for the relative loop suppression of $\cC_{e\gamma}$ compared to $\cC_{eH}$. We are estimating the future precision of $\Delta a_e$ to be $\Delta a_e^{\text{future}}\sim 5 \times 10^{-14}$.}
\end{table}

We can see from this table that if a model generates $\Delta \kappa_e$ and $\Delta a_e$ at the same loop order, then, given the current constraints on $\Delta a_e$, its contributions to electron Yukawa shifts are expected to be below $O(1)$ and hence unobservable at an FCC-ee Higgs pole run. This conclusion holds unless there is some fine-tuning or cancellation in the model which would alter this naive expectation. If instead a model generates $\Delta a_e$ with a 1-loop suppression relative to $\Delta \kappa_e$, then current constraints allow an $O(10)$---$O(100)$ deviation in the electron Yukawa, but future measurements of $\Delta a_e$ could reduce this to $O(1)$. In this case, which applies to the majority of the new states listed in Tab.~\ref{tab:UVcompletions}, the exact size of expected deviations in $\Delta \kappa_e$ will depend on specifics of the model, demanding a more complete analysis. Notice that many such states (specifically the ones inducing a tree level effect in $\cC_{eH}$) are highlighted in blue in Tab.~\ref{tab:UVcompletions}, signalling that the matching involves non-renormalisable interactions, or equivalently a UV completion involving an additional particle (or more), which can be from the same list above. Such two-particle models have been investigated in Refs.~\cite{Davoudiasl:2023huk,Erdelyi:2025axy}, finding that indeed future projections of $\Delta a_e$ can indirectly probe $\Delta \kappa_e$ generally to $O(1)$ levels in these models, but that in some instances projected bounds extend to $>10$ or $<1$. Instead, for the case where contributions to $\cC_{e\gamma}$ are suppressed by two additional loops as compared to $\cC_{eH}$, deviations in $\Delta \kappa_e$ could be very large, even if future measurements of $\Delta a_e$ are in agreement with SM predictions. 

In the following subsections, we investigate the two extremes of these possible loop suppressions in more detail. First, we explore the parameter space of the $\varphi$ model, including additional future tests of its couplings at FCC-ee and FCC-hh. Then, we look into models of scalar leptoquarks, exploring to what extent the link between $\Delta a_e$ and $\Delta \kappa_e$ can be altered in models with additional flavour structure.

\subsection{A heavy second Higgs doublet}

As argued above, the only tree level UV completion of $\Op_{eH}$ which gives a two loop relative suppression in $\Op_{e\gamma}$ consists of an extra scalar $SU(2)_L$ doublet with couplings to leptons and to Higgs bosons. This model hence represents the `best case scenario' for new physics in $\Delta\kappa_e$, in the sense that 
we might expect that current and near-future indirect constraints from $\Delta a_e$ could still allow very large deviations in $\Delta \kappa_e$. The possibility of a very large electron Yukawa within a 2HDM was studied previously in Ref.~\cite{Dery:2017axi}. Here, we take a simplified approach, assuming that $m_\varphi \gg m_h$, and only studying the minimal couplings needed to enhance the electron Yukawa coupling, to determine the impact of correlated tests of this scenario upcoming at HL-LHC or FCC-ee. The relevant Lagrangian is\footnote{Additional scalar potential terms would also be present, however we don't consider them here as they are not related directly to chirally enhanced $\kappa_e$ corrections.}
\begin{equation}
\label{eq:varphilag}
    -\mathcal{L}_\varphi \supset (y^e_{\varphi})_{ij} \varphi^\dagger\, \bar{e}_{R}^i\,l_{L}^j 
    + (y^{u}_{\varphi})_{ij} \, \varphi^\dagger i \sigma_2 \bar{q}_{Li}^T u_{Rj}
    +\lambda_\varphi\left(\varphi^\dagger H\right)\left(H^\dagger H\right) + \text{h.c}.
\end{equation}
Matching at tree level to the SMEFT gives \cite{deBlas:2017xtg}
\begin{align}
\label{eq:varphimatching}
    [\cC_{eH}]_{ij} = \frac{\lambda_{\varphi} ( y_{\varphi}^e)^*_{ji}}{M_{\varphi}^2}
    ,~
    [\cC_{le}]_{ijkl} = -\frac{(y_{\varphi}^e)^*_{li} (y_{\varphi}^e)_{kj} }{2 M_{\varphi}^2}
    ,~
    \cC_{\phi} = \frac{|\lambda_\varphi|^2}{M^2}
    ,~
    [\cC^{(1)}_{lequ}]_{ijkl} = -\frac{(y_{\varphi}^u)_{kl}(y_{\varphi}^e)^*_{ji}}{2 M_{\varphi}^2}.
\end{align}
Within this model, the Yukawa modifying coefficient, $\cC_{eH}$, is thus composed of two couplings which can be separately probed through other processes via contributions from $\cC_{\phi}$ and $\cC_{le}$.
Current direct constraints on $\cC_{le}$ are derived from Bhabha scattering measurements performed at LEPII \cite{Greljo:2024ytg,ALEPH:2013dgf,ALEPH:2010aa}. These constraints are expected to be significantly improved by off-$Z$-pole measurements of the leptonic ratio $R_e$ at FCC-ee \cite{Greljo:2024ytg}. Limits on $\cC_{\phi}$ can be placed from searches for double Higgs production at the LHC \cite{ATLAS:2022vkf,CMS:2022dwd} and HL-LHC \cite{deBlas:2944678}.  At FCC-ee the trilinear Higgs coupling can be probed at the $Zh$ run~\cite{McCullough:2013rea} while at FCC-hh measurements of double Higgs production will provide strong sensitivity \cite{FCC:2025lpp}.

\begin{figure}
    \centering
    \includegraphics[width=\linewidth]{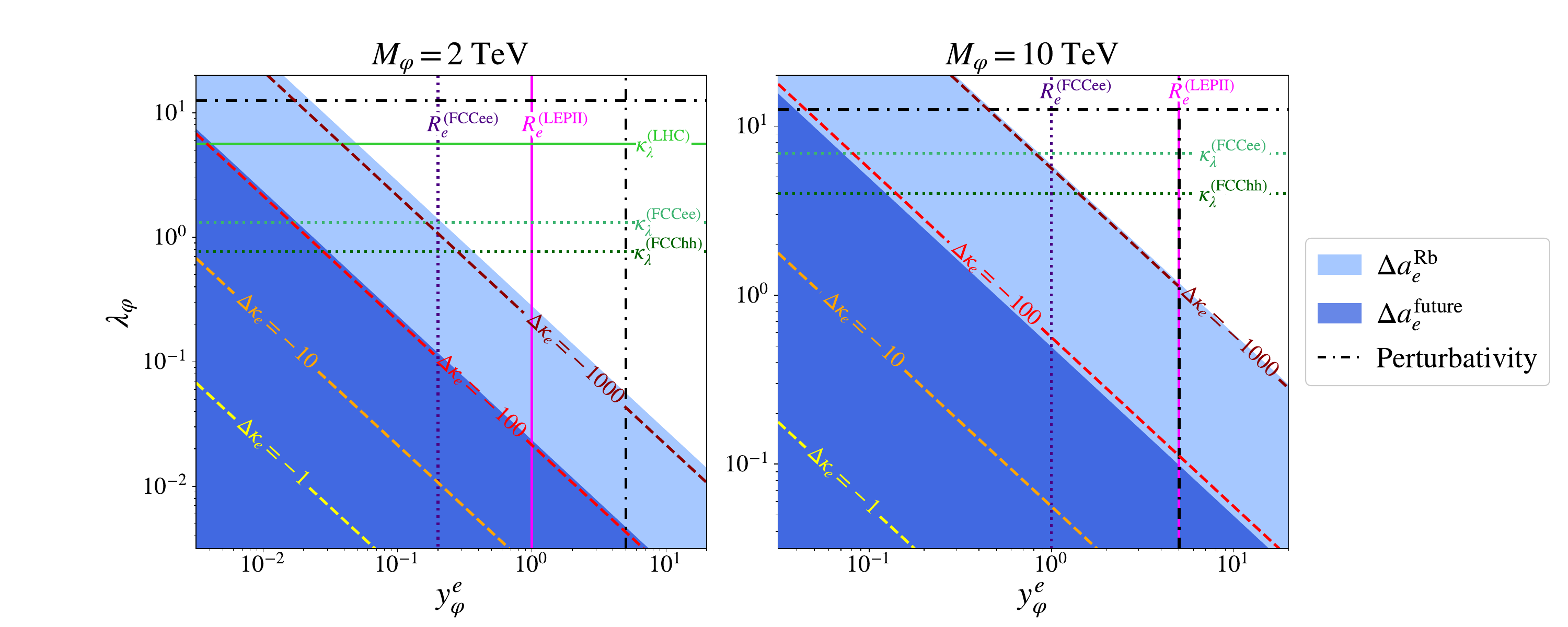}
    \caption{Parameter space for a $\varphi$ extension coupled to electrons. Blue shaded regions are consistent with the indicated $\Delta a_e$ measurement at 95\% C.L. Dashed lines show the $\kappa_e$ enhancement corresponding to the given parameter values. Solid (dotted) lines represent constraints on the couplings from current (projected) measurements at 95\% C.L. The $R_e$ constraints are taken from \cite{Greljo:2024ytg}, $\kappa^{\text{LHC}}_\lambda$ and $\kappa^{\text{FCCee}}_\lambda$ are from \cite{terHoeve:2025omu} and $\kappa^{\text{FCChh}}_\lambda$ assumes 6\%~uncertainty at 95\% CL \cite{FCC:2025lpp,deBlas:2944678}.}
    \label{fig:e_only_varphi}
\end{figure}

Figure~\ref{fig:e_only_varphi} shows these future projections in the plane of the two couplings $\lambda_\varphi$ and $(y^e_\varphi)_{11}$ for two different values of the mass, along with the current allowed region and future sensitivity of $\Delta a_e$. We assume here that only these two couplings are non-zero. Contours of different values of $\Delta \kappa_e$ are shown with dashed diagonal lines. It can be seen that current indirect constraints from $\Delta a_e$ allow extremely large values of $\Delta \kappa_e\sim 1000$, while the combination of measurements at FCC-ee and FCC-hh will indirectly probe values of couplings corresponding to $|\Delta \kappa_e| \lesssim 500$ at 2 $\sigma$. These are all larger than the current direct constraint on $\kappa_e$. Future measurements of $\Delta a_e$ could bring the indirect constraint on $\Delta \kappa_e$ down to about 100.

\begin{figure}
    \centering
    \includegraphics[width=1\linewidth]{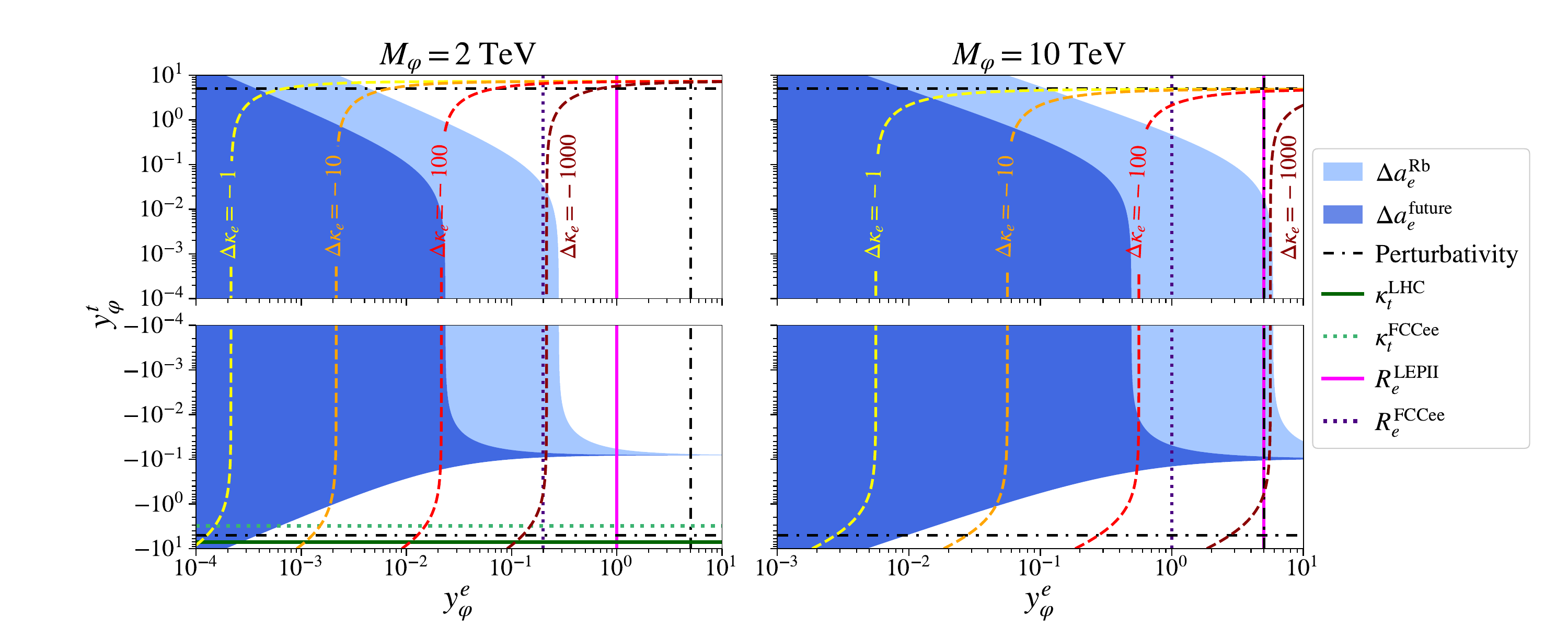}
    \caption{Parameter space for a second Higgs doublet $\varphi$ with Yukawa-like couplings to electrons and top quarks, and assuming $\lambda_\varphi = 1$. Shaded regions and line styles have the same meaning as in Fig.~\ref{fig:e_only_varphi}. The $\kappa_t^{\text{LHC}}$ limit is from \cite{ATLAS:2022vkf}, and the $\kappa_t^{\text{FCCee}}$ projection is from \cite{deBlas:2019rxi}. }
    \label{fig:e_t_varphi}
\end{figure}

As seen in \eqref{eq:varphimatching}, $\varphi$ also matches to $\cC^{(1)}_{lequ}$ at tree level if it couples to up type quarks. This coefficient generates contributions to $\Delta \kappa_e$ at the one loop level and to $\Delta a_e$ at two-loop level, with the largest effects from the operator involving the third generation quarks. From Fig.~\ref{fig:generating_ceh_from_running}, we see that this operator alone could induce $O(10)$ effects in $\Delta \kappa_e$ while remaining consistent with current constraints. To investigate this in the context of the $\varphi$ model, in addition to $\lambda_\varphi$ and $(y^e_\varphi)_{11}$, we now allow a non-zero value of $(y^u_\varphi)_{33}$. Fig. \ref{fig:e_t_varphi} shows the allowed regions for this scenario in the plane of $y_\varphi^e\equiv (y^e_\varphi)_{11}$ and $y_\varphi^t\equiv(y^u_\varphi)_{33}$, while holding the coupling to the Higgs boson fixed at $\lambda_\varphi=1$.

We see that including $O(1)$ Yukawa-like couplings to top quarks significantly tightens constraints from $\Delta a_e$ with respect to Fig.~\ref{fig:e_only_varphi}. However, at smaller values of $y_{\varphi}^t$ this effect quickly diminishes, until at $|y_{\varphi}^t|\sim 10^{-2}$, the one-loop contribution to $\Delta a_e$ (from $C_{lequ}^{1133}$) is subdominant to the two-loop contribution (from $C_{eH}^{11}$), and the situation becomes equivalent to the $y_{\varphi}^t=0$ case in Fig.~\ref{fig:e_only_varphi}. For negative $y_{\varphi}^t\sim -0.1$ a cancellation can occur between the $[\cC^{(1)}_{lequ}]_{1133}$ and $[\cC_{eH}]_{11}$ contributions to $\Delta a_e$ therefore lifting constraints. Such a cancellation would require significant tuning of the model parameters (including also the finite two-loop matching contributions at the UV scale which we have neglected in this analysis). 
Constraints on $y_{\varphi}^t$ can also be derived from measurements of the top quark Yukawa coupling modifier $\kappa_t$. However, the FCC-ee sensitivity to this coupling modifier does not improve on perturbativity bounds except in the negative quadrant of the $M_\varphi=2$~TeV plot. 

In summary, in a model with an extra Higgs doublet, there is open parameter space in which $|\kappa_e|\gg 1$ could arise, which could be tested by an FCC-ee Higgs pole run and which would not otherwise be probed in the near future by $\Delta a_e$ or by other FCC measurements.

\subsection{Leptoquarks}

Scalar and vector leptoquarks seemingly provide another promising option for new physics in $\kappa_e$, given they are the only extensions which match at tree level onto $\cC_{lequ}^{(1)}$ and $\cC_{ledq}$ respectively, both of which could naively produce large effects in $\kappa_e$ while remaining consistent with current constraints (see Fig.~\ref{fig:generating_ceh_from_running}). Inspection of Tab.~\ref{tab:UVcompletions}, however, reveals both the scalar leptoquarks ($\omega_1$ and $\Pi_7$) and the vector leptoquarks ($\mathcal{U}_2$ and $\mathcal{Q}_5$) actually correspond to the most pessimistic scenario for new physics, in which $\Delta \kappa_e$ and $\Delta a_e$ arise at the same loop order. For the scalar leptoquarks this is primarily due to a tree-level matching to $\cC_{lequ}^{(3)}$ \cite{deBlas:2017xtg} which runs at one loop into $\cC_{e\gamma}$, while for the vector leptoquarks the leading effect is a finite one-loop contribution to $\cC_{e\gamma}$ (see e.g.~\cite{Biggio:2016wyy, Queiroz:2014zfa, Altmannshofer:2020ywf}). Nevertheless, we explore the leptoquark option in more detail here, to examine whether these broad connections can be affected by particular parameter choices or flavour structures in the model.

We focus on the scalar leptoquark extensions $\omega_1$ and $\Pi_7$ where the relevant Lagrangian terms are
\begin{align}
-\mathcal{L}_{\text{LQs}} &\supset 
    (y^{ql}_{\omega_1})_{ij}
    \omega_{1}^\dagger \bar{q}^c_{Li} i\sigma_2 l_{Lj}
    +(y^{eu}_{\omega_1})_{ij}
    \omega_{1}^\dagger \bar{e}^c_{Ri} u_{Rj}
    +\lambda_{\omega_1} ( \omega_1^\dagger \omega_1 ) ( H^\dagger H )\nonumber\\
    &+(y^{lu}_{\Pi_7})_{ij} \Pi_{7}^\dagger i\sigma_2 \bar{l}^T_{Li} u_{Rj}
    + (y^{eq}_{\Pi_7})_{ij} \Pi_{7}^\dagger \bar{e}_{Ri} q_{Lj} \nonumber\\
    &+ \lambda_{\Pi_7} ( \Pi_7^\dagger \Pi_7 ) ( H^\dagger H )
+ \tilde{\lambda}_{\Pi_7} ( \Pi_7^\dagger H  ) ( H^\dagger \Pi_7 ) + \text{h.c},
\label{eqn:LQ_lagrangian}
\end{align}
where $i$ and $j$ are flavour indices, and colour and weak isospin indices are left implicit. These extensions match at tree level to the dimension-six SMEFT coefficients~\cite{deBlas:2017xtg}
\begin{align}
    \left(\mathcal{C}^{(1)}_{lequ}\right)_{ijkl} &= \frac{(y_{\omega_1}^{eu})_{jl}(y_{\omega_1}^{ql})^*_{ki}}{2M^2_{\omega_{1}}} + \frac{(y_{\Pi_7}^{eq})^*_{jk}(y_{\Pi_7}^{lu})_{il}}{2M^2_{\Pi_{7}}}, 
    \label{eqn:clequ1_matching}\\
    \left(\mathcal{C}^{(3)}_{lequ}\right)_{ijkl} &= -\frac{(y_{\omega_1}^{eu})_{jl}(y_{\omega_1}^{ql})^*_{ki}}{8M^2_{\omega_{1}}} + \frac{(y_{\Pi_7}^{eq})^*_{jk}(y_{\Pi_7}^{lu})_{il}}{8M^2_{\Pi_{7}}}.
    \label{eqn:clequ3_matching}
\end{align}
Since identical couplings appear in both matching expressions, the one-loop mixing of $\mathcal{C}^{(3)}_{lequ}$ into $\cC_{eB}$ and $\cC_{eW}$, and of $\mathcal{C}^{(1)}_{lequ}$ into $\cC_{eH}$, provides a strong connection between $\Delta \kappa_e$ and $\Delta a_e$ in these models.

We compute the one-loop matching of $\omega_1$ and $\Pi_7$ states onto all SMEFT operators contributing to $\Delta \kappa_e$ or $\Delta a_e$, with \texttt{SOLD}~\cite{Guedes:2023azv,Guedes:2024vuf} and provide these expressions in App.~\ref{app:lq_matching}. We neglect effects related to two-loop anomalous dimensions and two-loop matching at the EW scale.
The dependence on the up-type Yukawa, present identically in the RGEs of Eq.~\eqref{eq:ceh_rge} and the matching relations of App.~\ref{app:lq_matching}, indicates that leptoquark electron-top couplings ($(y_{\omega_1}^{eu})_{13}(y_{\omega_1}^{ql})_{31}^*$ or $(y_{\Pi_7}^{eq})_{13}(y_{\Pi_7}^{lu})_{13}^*$) will give dominant contributions to both $\Delta \kappa_e$ and $\Delta a_e$. Contributions from leptoquark electron-charm couplings ($(y_{\omega_1}^{eu})_{12}(y_{\omega_1}^{ql})_{21}^*$ or $(y_{\Pi_7}^{eq})_{12}(y_{\Pi_7}^{lu})_{12}^*$) are naively suppressed by at least $\frac{m_c}{m_t}$, but can in fact still be important due to RGE effects below the EW scale.
Within the matching relations, finite contributions proportional to quartic leptoquark couplings to the SM Higgs ($\lambda_{\omega_1}$ or $\lambda_{\Pi_7}$) are present but subdominant with respect to log-enhanced pieces. Nevertheless, these couplings can be phenomenologically relevant for Higgs-lepton coupling deviations \cite{Altmannshofer:2020ywf}, and we therefore include them in our analysis.

The ATLAS and CMS collaborations have searched for pair-produced scalar leptoquarks decaying to electron–top \cite{ATLAS:2024huc, ATLAS:2023prb, CMS:2022nty} and electron–charm \cite{ATLAS:2020dsk} final states, excluding masses up to $1.8$ TeV depending on coupling assumptions. We consider two benchmark scenarios, $M_{\omega_1/\Pi_7} = 2$ TeV and $10$ TeV, noting that the lighter case may be excluded by improved direct searches before a Higgs-pole run at FCC-ee. We perform tree- and one-loop matching to the SMEFT at these scales. The analysis then follows closely that of Sec.~\ref{sec:generating_ceh}; we compute the induced $\Delta \kappa_e$ and $\Delta a_e$ using the numerical solution for the one-loop RGEs from \texttt{DsixTools} and the results of \cite{Aebischer:2021uvt}. We provide expressions for these quantities in terms of the model parameters of Eq.~\ref{eqn:LQ_lagrangian}, at the benchmark mass scales, in App.~\ref{app:lq:expressions}.

\begin{figure}
    \centering
    \includegraphics[width=\linewidth]{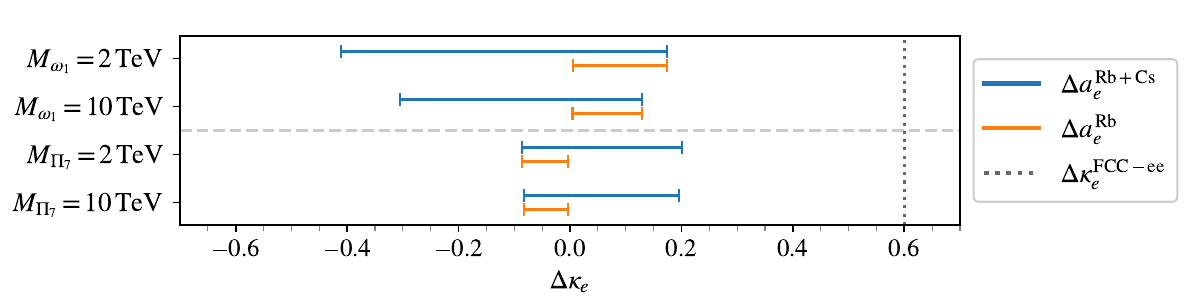}
    \caption{95\% C.L constraints on $\Delta \kappa_e$ implied by current measurements of $\Delta a_e$, assuming a UV completion of either $\omega_1$ or $\Pi_7$ with only electron-top couplings.}
    \label{fig:top_e_only}
\end{figure}

Current $\Delta a_e$ measurements already restrict $\Delta \kappa_e$ for $\omega_1$ and $\Pi_7$ with electron–top couplings, as shown in Fig.~\ref{fig:top_e_only}. Under either $\Delta a_e^{\text{Rb+Cs}}$ or $\Delta a_e^{\text{Rb}}$ assumptions, enhancements lie below projected FCC-ee sensitivity. These explicit bounds are slightly stronger than the model-independent estimates in Tab.\ref{tab:schematicboundsloops}.
For leptoquarks with only electron-charm couplings (previously studied as a way to explain deviations in lepton magnetic moments while avoiding strong $\mu \to e \gamma$ limits~\cite{Bigaran:2020jil, Dorsner:2020aaz}), we find that achieving $\Delta \kappa_e \sim 0.6$ would require $(y_{\omega_1}^{eu})_{12}(y_{\omega_1}^{ql})_{21}^*$ or $(y_{\Pi_7}^{eq})_{12}(y_{\Pi_7}^{lu})_{12}^*$ to be above the perturbativity bounds on this combination of couplings (see Eqs.~\eqref{eq:pi7_pert_limit} and \eqref{eq:omega1_pert_limit}). Hence, a single scalar leptoquark with only electron-charm couplings cannot generate large $\Delta \kappa_e$.

\begin{figure}
    \centering
    \includegraphics[width=1\linewidth]{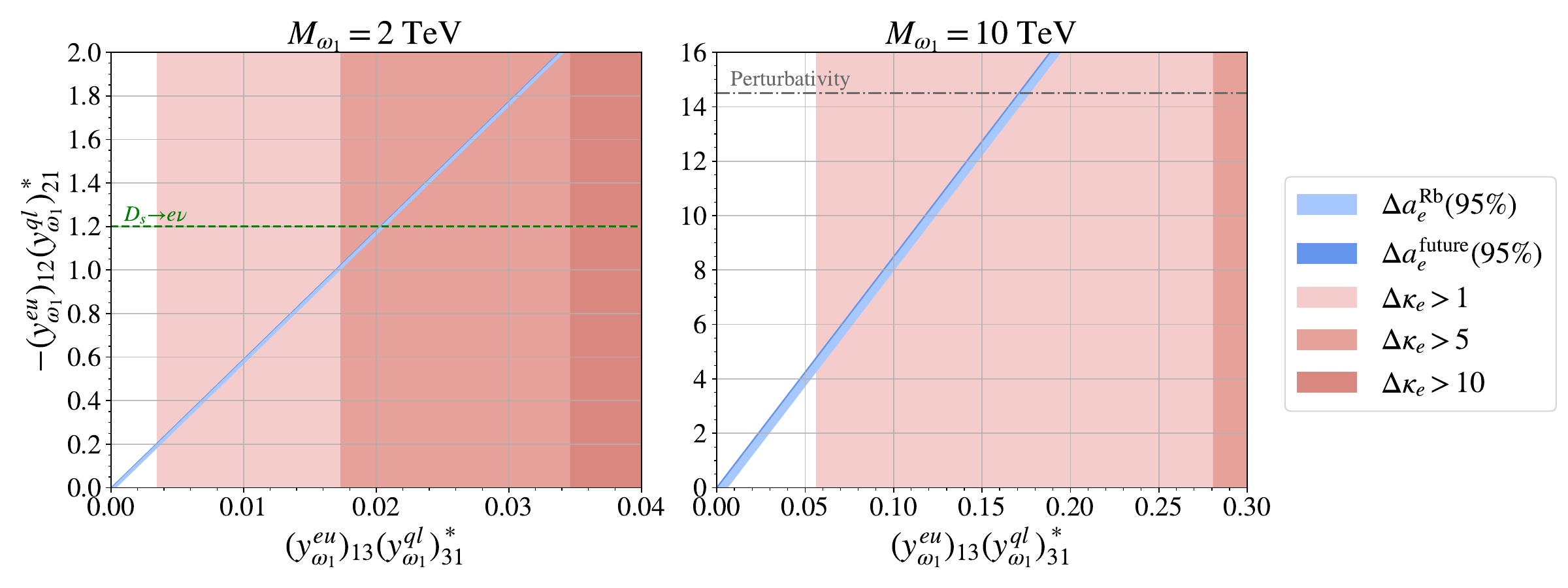}
    \caption{Parameter space for $\omega_1$ scalar leptoquarks with both electron-top and electron-charm couplings assuming $\lambda_{\omega_1} = 0$. Red and blue shaded regions are compatible with the stated $\Delta\kappa_e$ enhancement and measurement of $\Delta a_e$ respectively. The green dashed line shows the 95\% upper limit on the electron-charm couplings derived from current constraints on $\text{BR}(D^+_s\to e^+\nu)$. The darker blue region is not clearly visible in the left hand plot, but exists as a very narrow band at the top of the lighter blue band.}
    \label{fig:omega_1}
\end{figure}

Allowing both electron-top and electron-charm couplings, for specific coupling choices, cancellations can relax the strong constraints on $\Delta \kappa_e$ derived under the electron-top only assumption. 

We display the allowed parameter space for the $\omega_1$ extension in Fig.~\ref{fig:omega_1}. Here, the thinness of the blue bands demonstrates the current $\Delta a^{\text{Rb}}_e$ and $\Delta a^{\text{future}}_e$ mandate a significant tuning of the ratio between the couplings to generate a large $\Delta \kappa_e$\footnote{We note that the specific value of the required ratio depends weakly on the non-perturbative parameter $c_T^{(c)}$ \cite{Aebischer:2021uvt}. The dependence of $\Delta a_e$ on this parameter is given in App.~\ref{app:lq:expressions}. We set this to zero here, but an $O(1)$ value for this parameter would change the required coupling ratio by $O(10\%)$.}.
For $M_{\omega_1}=2$~TeV, tuning can allow $\Delta \kappa_e \sim -70$, whilst not violating perturbativity bounds on electron-charm couplings, however constraints from $D_s\to e \nu$ (discussed in App.~\ref{sec:mesondecays}) restrict this maximum enhancement to $\Delta \kappa_e \sim -6$. A heavier $M_{\omega_1}=10$~TeV field will evade any meson decay or pair-production bounds but has a lower maximum enhancement magnitude $\Delta \kappa_e\sim-3$. Both cases represent parameter space which will be tested by an FCC-ee $\kappa_e$ measurement at a dedicated Higgs pole run.  Similar conclusions hold for the $\Pi_7$ state when both electron-top and electron-charm couplings are allowed. We provide the analogous plots for this state in App.~\ref{app:pi7_plots}. 

Such a situation, with both top-electron and top-charm couplings, would imply that the leptoquarks would also mediate flavour changing neutral currents involving the top. Specifically, decays $t\to c \,e^+ e^-$ are generated at tree level through some combinations of these couplings, while $t\to c Z$ would be induced at one loop by closing the electron loop and radiating a $Z$. Current constraints are only set on the latter process and do not effectively bound the model at all, see App.~\ref{sec:topfcncs}. Projected $t\to cZ$ constraints at FCC-ee and FCC-hh are also very weak for this model. Given that it occurs at tree level, a promising search channel at future $e^+e^-$ colliders is $e^+e^-\to t \bar{c}$. Corresponding projected constraints for a 240 GeV run at CEPC~\cite{Shi:2019epw} are at the level of $[y^{ul}]_{21}[y^{qe}]_{31}<0.6$ for a leptoquark of mass $2$ TeV (see App.~\ref{sec:topfcncs}), and could provide a test of the model, depending on the leptoquark masses that remain viable after HL-LHC. However we note that this is sensitive to a different combination of couplings to that appearing on the axes of Fig.~\ref{fig:omega_1}, so this projection cannot be shown directly on this plane.

\begin{figure}
    \centering
    \begin{subfigure}[t]{0.48\linewidth}
        \centering
        \includegraphics[width=\linewidth]{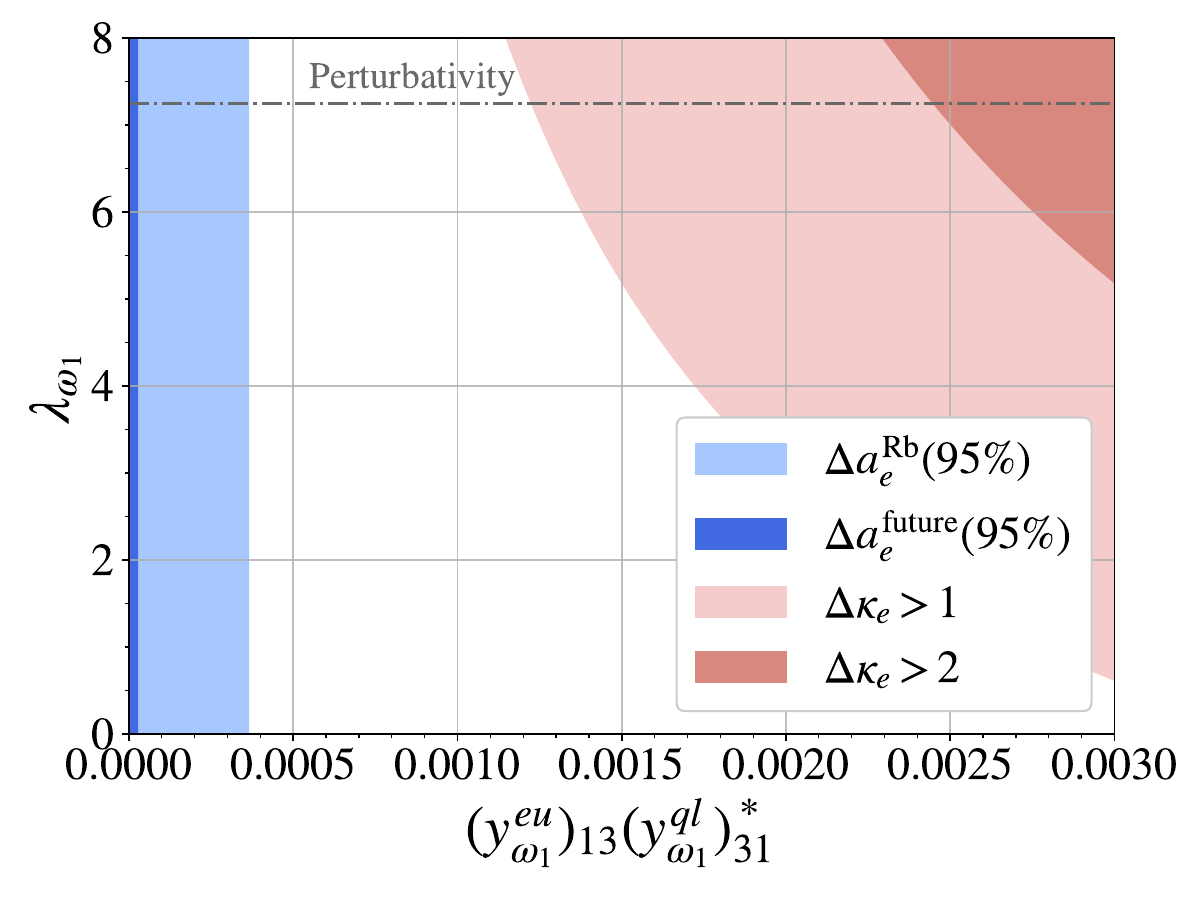}
        \caption{}
        \label{fig:omega1_lambda_a}
    \end{subfigure}
    \hfill
    \begin{subfigure}[t]{0.48\linewidth}
        \centering
        \includegraphics[width=\linewidth]{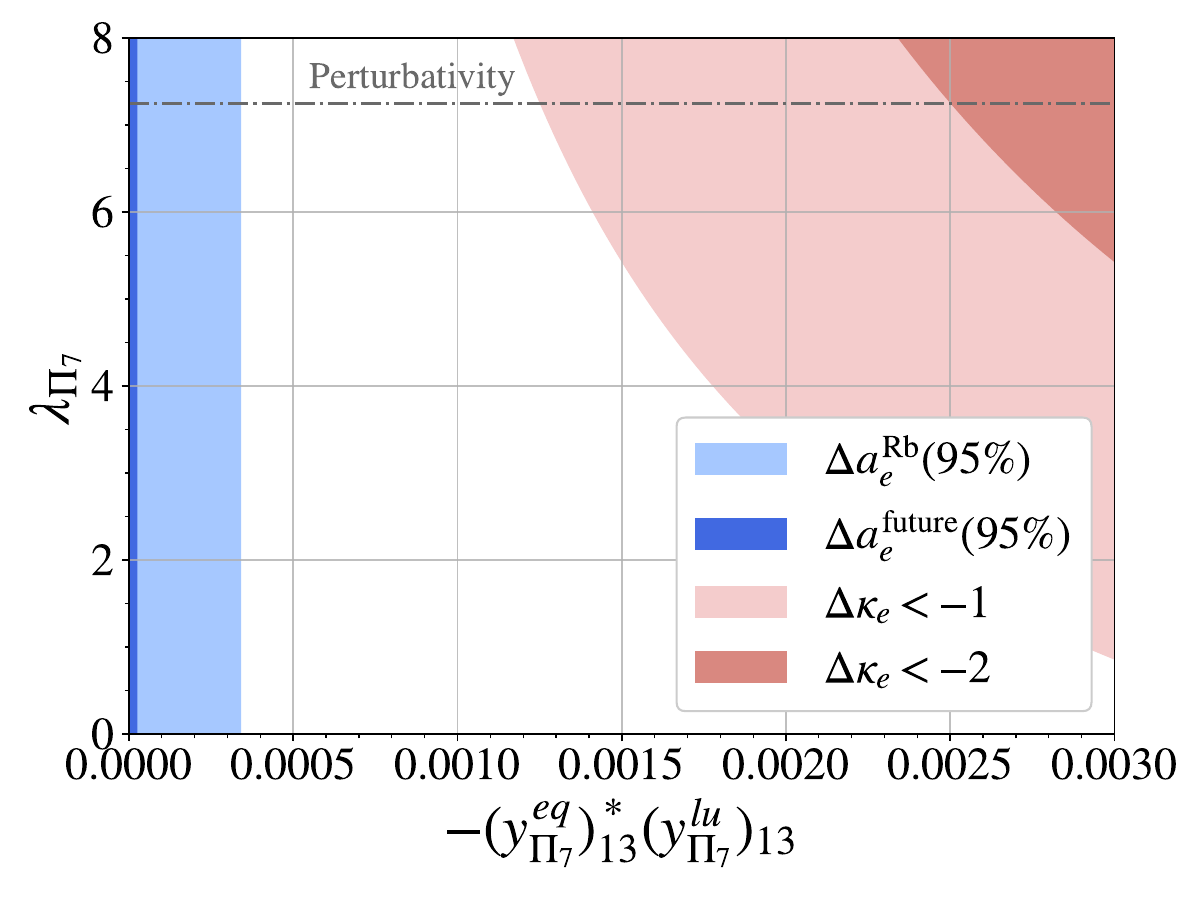}
        \caption{}
        \label{ig:omega1_lambda_b}
    \end{subfigure}
    \caption{Parameter space for $\omega_1$ (a) and $\Pi_7$ (b) leptoquarks assuming $M_{\omega_1/\Pi_7}=2$ TeV and only $(y_{\omega_1}^{eu})_{13}(y_{\omega_1}^{ql})_{31}^*$ and $\lambda_{\omega_1}$ or $(y_{\Pi_7}^{eq})_{13}(y_{\Pi_7}^{lu})_{13}^*$ and $\lambda_{\Pi_7}$ are non-zero.}
    \label{fig:leptoquarks_lambda_plots}
\end{figure}

Finally, we examine whether cancellations between running contributions to $\Delta a_e$ and finite effects from the quartic couplings $\lambda_{\omega_1}$ and $\lambda_{\Pi_7}$ can allow for large $\Delta \kappa_e$. Neglecting electron–charm couplings now, Fig.~\ref{fig:leptoquarks_lambda_plots} shows that for $M_{\omega_1/\Pi_7}=2$~TeV neither state can achieve $\Delta \kappa_e > 0.6$ whilst respecting $\Delta a_e$ and perturbativity constraints. For heavier masses, the allowed $\Delta \kappa_e$ enhancements are further diminished.

In summary, scalar leptoquarks are unlikely to contribute to a measurable deviation in $\kappa_e$ at an FCC-ee Higgs pole run, since the corresponding couplings are already well tested by $\Delta a_e$. The only way that a significant deviation in $\kappa_e$ can be achieved is by a leptoquark with a tuned ratio of charm-electron to top-electron couplings, and in which the charm-electron coupling is significantly (about 60 to 80 times) larger than the top-electron coupling, and of opposite sign. This is a rather unnatural flavour structure, which would appear to have no justification other than to achieve this effect.

\subsection{Comments on fine-tuning}
\label{sec:radiative}
If the leading correction arises at dimension-6, then to realise a large ($\gg 1$) Yukawa modification requires fine-tuning the dimension-4 Yukawa coupling such that the electron mass is small despite having a large coupling to the Higgs. The greater the fine-tuning, the greater the coupling enhancement.  Indeed, one may quantify this for the dimension-6 contribution in the usual way, through a fine-tuning estimate
\begin{equation}
    \Delta = \left|\frac{\partial \ln \kappa_e}{\partial \ln y_e}\right| = \left|\frac{(1-\kappa_e)(3-\kappa_e)}{2 \kappa_e} \right|  ~~,
\end{equation}
which behaves as
\begin{equation}
    \lim_{\kappa \gg1} \Delta = \frac{\kappa_e }{2} ~~,
\end{equation}
and exhibits vanishing fine-tuning for $\kappa_e = 1$ or $\kappa_e = 3$, reflecting the cases where the dim-4 or dim-6 contributions dominate.

As a result, we see that in the limit where a new physics scenario gives rise to a Yukawa coupling which is significantly greater in magnitude than the SM coupling and not an integer (which corresponds to a single operator dominating), the parameters \emph{within the UV theory} will have to be fine-tuned by an amount directly proportional to the magnitude of the enhancement. Pragmatically speaking, a theory which enhances the electron Yukawa by a factor of $10$ will generically need to be fine-tuned at around the $20\%$ level. 
Realistic UV-completions give rise to a tower of operators, thus this dimension-6 based analysis will be quantitatively modified.  Nonetheless, the qualitative picture will remain valid as having a large Higgs coupling but small mass will generically require fine-tuning to realise the latter.

One should also consider radiative corrections from an IR perspective.  Na\"ively there is no symmetry nor dimensional analysis of Eq.~\eqref{eq:Yuk6} which can, from an IR perspective, forbid corrections at the matching scale or within the full UV theory of the form
\begin{equation}
    \delta \mathcal{L}_{\text{Yuk}} = \frac{1}{(4 \pi)^2} M^2 C_{eH} H \bar l e +... ~~,
\end{equation}
where $M$ is a UV mass scale.  That this does not overcome the ultimate Yukawa correction originating from the dimension-6 contribution requires that
\begin{equation}
    v^2 \gtrsim \frac{M^2}{(4 \pi)^2}  ~~,
\end{equation}
na\"ively requiring that any new states ought not to be heavier than $\sim 3$ TeV.\footnote{This aspect is alluded to in \cite{Giudice:2008uua} and has independently been emphasised to MM by Neal Weiner and Riccardo Rattazzi.}  This has little impact on the models considered here as it stands, however it does suggest that if a significant Yukawa modification were observed then natural new physics shouldn't lie too far above the TeV scale, despite that the electron Yukawa coupling is so small.  However, as the previous discussion reveals, if $> \mathcal{O}(1)$ Yukawa coupling modifications are realised then some fine-tuning between higher and lower dimension contributions is necessary in any case, so naturalness arguments are weakened and it could be that the new physics responsible lies at higher mass scales.

We now turn to the question of RGE effects \emph{within} the UV theory.  One expects, on symmetry grounds, that within the UV theory there will be RGE contributions to $y_e$ proportional to $C_{eH}$.  Being model-dependent, it is not a theorem that such contributions must exist, however this is an important consideration since it impacts the question of radiative stability and fine-tuning of parameters within the UV theory.  The reason is that we have already established that the value of $y_e$ at the matching scale must be fine-tuned in order to realise significantly enhanced couplings. 
This fine-tuning is exacerbated if the precise value of $y_e$ is not radiatively stable within the UV theory, which in the absence of additional symmetries is not generically expected. If the UV theory does appear radiatively stable, one should scrutinise it to ascertain if some fine-tuning, including RG effects, has been employed elsewhere in order to render small the couplings which contribute to the running of $y_e$. 

\section{Conclusions}
\label{sec:conc}
The electron Yukawa is presently very poorly constrained.
Higgs boson decay measurements are projected, at HL-LHC, to be sensitive to up to 120 times the SM value.
This could be improved on significantly during a proposed FCC-ee run on the Higgs resonance, in which it can be accessed through $s$-channel production.
Although it is not part of the baseline running scenarios as outlined in the recently published FCC feasibility study \cite{FCC:2025lpp}, if it goes ahead this run would test electron Yukawa couplings down to 1.6 times the SM value.
As a fundamental Standard Model parameter, confirming the value of this coupling would be an important milestone, but it is worth evaluating what any deviation in this coupling would tell us about BSM physics.
In this work we have explored the broad theoretical implications of an $\gtrsim\mathcal{O}(1)$ enhancement of the electron Yukawa, correlating it with other measurements in the landscape of current and future planned experiments.

In the EFT language, a large enhancement in $\kappa_e$ can only be achieved by the operator $[\mathcal{O}_{eH}]_{11}$. This is one of only six Warsaw basis operators which break electron chiral symmetry. We have focussed on these operators in the first part of this work, noting in particular that a non-zero coefficient for $\mathcal{O}_{eH}$ generates a contribution to the electron electromagnetic dipole operator at two loops.
Estimates for the future precision on the electron magnetic dipole moment $\Delta a_e$ would hence yield an indirect $95\%$ bound on the coupling modifier $\kappa_e$ of approximately $70$ -- $90$, depending on the scale of new physics.

A contribution to $[\mathcal{O}_{eH}]_{11}$ could in turn be generated by RG running from another of the operators which break electron chiral symmetry. By examining current constraints on these operators from Drell-Yan processes at the LHC and $\Delta a_e$, we determined that new physics which generates (one of) the coefficients $[C_{lequ}^{(1)}]_{1133}$ or $[C_{ledq}]_{1133}$ could be consistent with current constraints while inducing an enhanced value of $\kappa_e$ detectable at a dedicated FCC-ee run. The relevant parameter space for these scenarios would, however, be independently tested by improved measurements of $\Delta a_e$ and hadronic ratios above the Z-pole.

Another important aspect concerns the UV flavour structure of the $O_{eH}$ operator.
In a flavour-anarchic scenario, large effects are to be expected both in flavour-violating Higgs decays and in $\mu\to e$ conversion.
Current experimental constraints show compatibility with the anarchic scenario only for $|\Delta \kappa_e| \lesssim 1$, already in the ballpark of FCC-ee projections.  Future constraints could indirectly constrain BSM contributions much more strongly, for instance with COMET \cite{Moritsu:2022lem}, reaching the $|\Delta \kappa_e| \lesssim 0.05$ level before FCC-ee operation.
A flavour-universal aligned NP scenario would allow for $|\kappa_e| \lesssim 10$, given the precision on the muon Yukawa expected from HL-LHC.  However, the anticipated precision on the muon Yukawa at FCC-hh would indirectly constrain $|\Delta \kappa_e| \lesssim 1$, under this ansatz, were the FCC-ee Higgs pole run not to have taken place.  An MFV scenario gives equal coupling modifications across the generations and thus an $\mathcal{O}(1)$ electron Yukawa modification is already ruled out. The only flavour scenario in which the Higgs-pole run at FCC-ee could provide the strongest sensitivity compared to other future tests is an electrophilic one.

In the second part of this paper, we studied explicit simplified models, contrasting their general features with the above EFT-based conclusions. In particular, we argued that within explicit models, the electromagnetic dipole operator $\mathcal{O}_{e\gamma}$ is generically expected to arise with only a one-loop suppression relative to the $\mathcal{O}_{eH}$ operator. This tightens the corresponding indirect constraints on $\kappa_e$ from $\Delta a_e$, meaning that estimates for future $\Delta a_e$ precision are expected to be sensitive to $O(1)$ deviations in $\kappa_e$. Thus, depending on improvements in $\Delta a_e$ by the time of FCC-ee, the parameter space that would be probed by an improved limit on $\kappa_e$ may already be ruled out in most models. The one important exception is a model with a second Higgs doublet, which can generate $\kappa_e$ at tree level, but $\Delta a_e$ only at two loops. Much parameter space of a simplified model of a heavy Higgs doublet would remain untested by other measurements, allowing large $\kappa_e$ deviations.

The option of generating $\kappa_e$ at loop level from $[C_{lequ}^{(1)}]_{1133}$ or $[C_{ledq}]_{1133}$, which appeared viable within the EFT picture, implies a leptoquark UV completion. We studied simple models of scalar leptoquarks, noting that all generate $\Delta a_e$ at the same loop order as $\kappa_e$, and hence are not expected to produce significant electron Yukawa deviations given current constraints. For scalar leptoquarks with both electron-top and electron-charm couplings, large values of $\kappa_e$ (in the $\mathcal{O}(10)$ range) can be compatible with current and future constraints, but at the price of a significant fine-tuning in flavour space.

Finally, we note that any large Yukawa modification necessarily comes with some fine-tuning between the dimension-four and dimension-six contributions.
On top of that, considerations about radiative stability due to contributions of the dimension-six term in the running of the renormalisable Yukawa lead to the conclusion that the new states involved should have mass satisfying $M \lesssim 4\pi v$.

In conclusion, an $\mathcal{O}(1)$ modification of the electron Yukawa can be consistent with reasonable assumptions about perturbativity and fine-tuning, but only if new physics has a very different flavour structure to the SM.
We have identified the conditions on the UV for which a direct measurement can be the most relevant constraint. Models of new physics which match directly to the $\mathcal{O}_{eH}$ operator, and which couple most strongly to the electron, could be best tested here. 
However, the tight connections between the Yukawa operators and the dipole operators means that generically a deviation in the electron Yukawa implies a deviation in $\Delta a_e$. The added value of a Higgs pole run at FCC-ee for our understanding of BSM physics will hence partly depend on future improvements in the sensitivity on the anomalous magnetic moment of the electron.

\acknowledgments
We are grateful to Jorge de Blas, Gauthier Durieux, Guilherme Guedes and Riccardo Rattazzi for discussions, and to Marc Riembau for correspondence concerning Ref.~\cite{Panico:2018hal}.
SR is supported by UKRI Stephen Hawking Fellowship EP/W005433/1, and partially supported by STFC grant ST/X000605/1. LA acknowledges funding from the Deutsche Forschungsgemeinschaft under Germany’s Excellence Strategy EXC 2121 “Quantum Universe” – 390833306, as well as from the grant 491245950. BS is supported by a Lord Kelvin Adam Smith scholarship from the University of Glasgow. This research was supported in part by grant NSF PHY-2309135 to the Kavli Institute for Theoretical Physics (KITP).

\appendix

\section{Theoretical expressions for meson decays}
\label{sec:mesondecays}
The EFT operator coefficients $\cC_{ledq}$ and $\cC_{lequ}^{(1,3)}$ can contribute to electronic meson decays with large chiral enhancements over the SM.

\subsection{$B_s\to ee$}
The low-energy effective Lagrangian describing $b\to s \ell\ell$ transitions is given by
\begin{equation}
    \mathcal{L}_{bs\ell\ell}\supset \frac{4G_F}{\sqrt{2}}V_{ts}^*V_{tb}\sum_k C_k O_k,
\end{equation}
with operators 
\begin{align}
    O_9&=\frac{\alpha}{4\pi}(\bar{s}_L\gamma_\mu b_L)(\bar{\ell}_i\gamma^\mu\ell_i), ~~~~O_{10}=\frac{\alpha}{4\pi}(\bar{s}_L\gamma_\mu b_L)(\bar{\ell}_i\gamma^\mu\gamma^5\ell_i),\\
    O_S&=\frac{\alpha}{4\pi}(\bar{s}_L b_R)(\bar{\ell}_i\ell_i), ~~~~~~~~~~O_P=\frac{\alpha}{4\pi}(\bar{s}_L b_R)(\bar{\ell}_i\gamma^5\ell_i).
\end{align}
The SM values of the corresponding Wilson coefficients are \cite{Allwicher:2025bub}:
\begin{align}
    C_{9}^{\text{SM}}=4.114, ~~C_{10}^{\text{SM}}=-4.193, ~~C_S^{\text{SM}}=0, ~~C_P^{\text{SM}}=0.
\end{align}
The SMEFT Wilson coefficient $\cC_{ledq}$ with third generation quark indices matches at the electroweak scale onto the coefficients $C_S$ and $C_P$ as:
\begin{equation}
    C_S=-C_P=\frac{\pi v^2}{\alpha V_{tb}}[\cC_{ledq}]_{ii33},
\end{equation}
where the quark doublet flavour index is defined in the \emph{up quark mass basis}, i.e.~$q_3\equiv (t, V_{tj} d_j)^T$. Note that if instead this coefficient were aligned in the down quark mass basis, there would be no down-type flavour change.

The coefficients $C_S$ and $C_P$ receive a multiplicative factor from the running from the electroweak scale to the $b$ mass scale: $C_{S,P}(m_b)=1.38\,C_{S,P}(m_Z)$ \cite{Jenkins:2017jig}. Then the branching ratio for $B_s\to ee$ is given by~\cite{Allwicher:2025bub}:
\begin{align}
    \text{BR}(B_s\to ee)&=\text{BR}(B_s\to ee)_{\text{SM}}\left(\left| 1+1.38\frac{C_P}{C_{10}^{\text{SM}}} \frac{m_{B_s}^2}{2m_e(m_b+m_s)}\right|^2\right.\nonumber\\&+ \left.\left|1.38\frac{C_S}{C_{10}^{\text{SM}}} \frac{m_{B_s}^2}{2m_e(m_b+m_s)}\right|^2\right).
\end{align}
The factor of $m_e$ in the denominator of the terms dependent on $C_S$ and $C_P$ gives an enormous chiral enhancement to decays mediated by these scalar and pseudoscalar operators. The SM prediction for this decay is very small \cite{Bobeth:2013uxa,Beneke:2019slt}:
\begin{equation}
    \text{BR}(B_s\to ee)_{\text{SM}}=(8.30\pm 0.36)\times 10^{-14},
\end{equation}
and experimentally only an upper limit has been set, several orders of magnitude larger than the SM expectation~\cite{LHCb:2020pcv}:
\begin{equation}
    \text{BR}(B_s\to ee) < 11.2\times 10^{-9}\,~ (95\% \,\text{CL}).
\end{equation}
This translates to a bound on $[\cC_{ledq}]_{1133}$ of
\begin{equation}
   | [\cC_{ledq}]_{1133}|<4.1\times 10^{-3} \, \text{TeV}^{-2}\,~ (95\% \,\text{CL}).
\end{equation}
We emphasise that the flavour index on the quark doublet is assumed to be in the up-aligned basis here. If instead the index is down-aligned, then this coefficient does not mediate a $b\to s$ flavour change and there is no bound on the coefficient from this process.

\subsection{$D_s\to e \nu$}
The low-energy effective Lagrangian describing $c\to s \ell \nu$ decays is written
\begin{align}
    \mathcal{L}_{cs\ell\nu}&=-\frac{4G_FV_{cs}}{\sqrt{2}}\bigg[(1+\epsilon_L)(\bar c_L\gamma^\mu s_L)(\bar \ell_L\gamma_\mu \nu_L)+ \epsilon_R(\bar c_R\gamma^\mu s_R)(\bar \ell_L\gamma_\mu \nu_L) \nonumber\\
    &+\epsilon_T(\bar c_L\sigma^{\mu\nu} s_L)(\bar \ell_L\sigma_{\mu\nu} \nu_L)+\epsilon_{S_L}(\bar c_R s_L)(\bar \ell_R\nu_L)+\epsilon_{S_R}(\bar c_L s_R)(\bar \ell_R\nu_L)\bigg]+h.c..
\end{align}
In the SM, all the $\epsilon$ coefficients are zero. The branching ratio of $D_s^+\to e^+ \nu$ is given by \cite{Gonzalez-Alonso:2016etj}
\begin{equation}
    \text{BR}(D_s^+\to e^+ \nu)=\frac{G_F^2}{8\pi}\tau_{D_s}f_{D_s}^2 |V_{cb}|^2 m_{D_s}m_e^2\left|1+\epsilon_L+\frac{m_{D_s}}{m_e(m_c+m_s)}(\epsilon_{S_R}-\epsilon_{S_L})\right|^2.
\end{equation}
where the $\epsilon$ coefficients are asssumed to be defined at the $c$ mass scale. The decay constant is $f_{D_s}=249.9\pm 0.5$ MeV~\cite{Carrasco:2014poa,Bazavov:2017lyh,FlavourLatticeAveragingGroupFLAG:2024oxs}. The SMEFT coefficient $[\cC_{lequ}^{(1)}]_{1122}$ matches to $\epsilon_{S_L}$ as:\footnote{Note that in general, there are other SMEFT matching relations that could in principle play a role; for example $[\cC_{lequ}^{(3)}]_{1122}$ matches to $\epsilon_{T}$ and $[\cC_{ledq}]_{1122}$ matches to $\epsilon_{S_R}$. However the former does not contribute to the $\text{BR}(D_s^-\to e^- \bar\nu)$ (except in very small amounts through RG mixing of $\epsilon_{T}$ into $\epsilon_{S_L}$), and the latter does not feature in our main analysis since it cannot contribute measurably to the electron Yukawa.}
\begin{align}
    \epsilon_{S_L} (m_c)= -1.78 \frac{v^2}{2V_{cs}}[\cC_{lequ}^{(1)*}]_{1122}(m_W),
\end{align}
where the factor of $1.78$ arises from the running from the electroweak scale to the $c$ mass scale \cite{Jenkins:2017jig}. We assume here that the quark doublet flavour index is defined in the up mass basis; if instead it is defined in the down mass basis, the factor of $V_{cs}$ would not appear in the denominator of the matching expression. 

The measured upper bound on the branching ratio is three orders of magnitude larger than the SM expectation \cite{Belle:2013isi}
\begin{equation}
     \text{BR}(D_s^+\to e^+ \nu)<1.0\times 10^{-4}\,~(95\% \,\text{CL}),
\end{equation}
leading to a bound on the SMEFT Wilson coefficient of 
\begin{align}
  |[\cC_{lequ}^{(1)*}]_{1122}|<0.18 \, \text{TeV}^{-2}\,~ (95\% \,\text{CL}).
\end{align}

\section{$\mu \to e$ Conversion} \label{app:mu2eConversion}
Assuming all of the quark Yukawas take their SM values, the leading contributions to the rate for $\mu \to e$ conversion can be written following Appendix~2 in \cite{Harnik:2012pb},
\begin{align}
    \Gamma(\mu \to e) = & \left| -\frac{D}{2m_\mu} [L_{e\gamma} (m_h)]_{e\mu} + \tilde g_{LS}^{(p)} S^{(p)} + \tilde g_{LS}^{(n)} S^{(n)} + \tilde g_{LV}^{(p)} V^{(p)} \right|^2 \  + \\
    & \left| -\frac{D}{2m_\mu} [L_{e\gamma} (m_h)]_{\mu e}^* + \tilde g_{RS}^{(p)} S^{(p)} + \tilde g_{RS}^{(n)} S^{(n)} + \tilde g_{RV}^{(p)} V^{(p)} \right|^2 ~, 
\end{align}
where the terms containing $L_{e\mu}$ follow from photon-exchange t-channel diagrams, the $S$ terms follow from tree-level Higgs exchange, and the $V$ terms from an evaluation of the 1-loop diagram. For further detail on the $V$ term, we refer to ref.~\cite{Harnik:2012pb}, however assuming $\cC_{eH}$ is real, we find approximately that
\begin{align}
    \tilde g_{LV/RV} \approx (1.1 \times10^{-9}) \times  \frac{\alpha v^2}{2\sqrt 2\pi m_\mu^2}  [\cC_{eH}(m_h)]_{e\mu/\mu e}.
\end{align}
The $S$ terms contain quark Yukawas and factors of the nuclear matrix elements $f^{(q,N)} = \bra{N} m_q \bar qq \ket{N} / m_p$ where $N \in \{n,p\}$. In our scenario they are given by
\begin{align}
    g_{LS}^{(N)} &=  -\sqrt 2 \frac{m_p v}{m_h^2} [\cC_{eH}(m_h)]_{e\mu} \sum_q f^{(q,N)}
    \\
    g_{RS}^{(N)} &=  -\sqrt 2 \frac{m_p v}{m_h^2} [\cC_{eH}(m_h)]_{\mu e}^* \sum_q f^{(q,N)} .
\end{align}
Numerically, the matrix elements $f^{(q,N)}$ are given as well in ref.~\cite{Harnik:2012pb}, taking the values
\begin{align}
    &f^{(u,p)} = f^{(d,n)} = 0.024 ~~~,~~~ f^{(d,p)} = f^{(u,n)} = 0.033 
    \\ &f^{(s,p)} = f^{(s,n)} = 0.25
    ~~~~~\,,~~~ f^{(other,p)} = f^{(other,n)} = 0.051 \ .
\end{align}
Finally, in units of $m_\mu^{5/2}$, nuclear overlap integrals and muon capture rates are given by ref.~\cite{Kitano:2002mt}, summarized in the following table.
\begin{table}[h]
\centering
\begin{tabular}{|c|c c c c c|}
\hline 
     & $D$ & $S^{(p)}$ & $S^{(n)}$ & $V^{(p)}$ & $\Gamma_{\rm capture}/(10^6 s^{-1})$\\ \hline
    Gold & 0.189 & 0.0614 & 0.0918 & 0.0974 & 13.07\\
    Aluminum & 0.0362 & 0.0155 & 0.0167 & 0.0161 & 0.7054\\
    \hline
\end{tabular}
\caption{Wavefunction overlap integrals for muons captured in the two relevant nuclei.}
\end{table}

\section{Perturbativity bounds}
Following the general discussion in \cite{Allwicher:2021rtd}, we report here the perturbativity limits on relevant coupling combinations for the models we study in Sec.~\ref{sec:models}.
For the $\Pi_7$ leptoquark, we follow the discussion in \cite{Allwicher:2021jkr}, which gives
\begin{align}
    \Re[y_{\Pi_7}^{ul} y_{\Pi_7}^{qe*}] \lesssim \frac{8\pi}{\sqrt{3}} \simeq 14.5 \qquad \lambda_{\Pi_7} \lesssim \frac{4\pi}{\sqrt{3}} \,.
\label{eq:pi7_pert_limit}
\end{align}
Similarly, the case of the $\omega_1$ leptoquark gives
\begin{align}
    \Re[y_{\omega_1}^{eu}y_{\omega_1}^{ql*}] \lesssim \frac{8\pi}{\sqrt{3}} \qquad \lambda_{\omega_1} \lesssim \frac{8\pi}{\sqrt{6}} \,.
\label{eq:omega1_pert_limit}
\end{align}
For the $\varphi$ couplings, partial-wave unitarity yields
\begin{align}
    \lambda_\varphi \lesssim 4 \pi \qquad y_e^\varphi \lesssim \sqrt{8\pi} \,.
\end{align}

\section{More details on scalar leptoquarks}

\subsection{Matching expressions for scalar leptoquarks}
\label{app:lq_matching}
We provide here the expressions for the one loop matching of $\omega_1$ and $\Pi_7$ onto relevant SMEFT coefficients, 
\begin{align}
    \left[\cC_{eH}\right]_{ij}=&\frac{3}{16\pi^2}\left[\frac{(y_{\omega _1}^{eu})_{jp}
   (y_{\omega _1}^{ql})^*_{ri}}{M^2_{\omega_1}}[Y_u^\dagger Y_u Y_u^\dagger]_{rp}\left(1+\ln\left(\frac{\mu ^2}{M_{\omega _1}^2}\right)\right) \right. \nonumber\\
   &\left. -\frac{(y_{\omega _1}^{eu})_{j p} (y_{\omega _1}^{ql})_{ri}^{* }}{2M_{\omega_1}^2}[Y_u^{\dagger}]_{rp} \left(2 \lambda _{\omega _1}+\lambda _{\phi } \left(3 + 2 \ln \left(\frac{\mu ^2}{M_{\omega _1}^2}\right)\right)\right) \right. \nonumber\\
   &\left. +\frac{(y_{\Pi_7}^{eq})^*_{jp}
   (y_{\Pi _7}^{lu})_{ir}}{M_{\Pi_7}^2}[Y_u^\dagger Y_u Y_u^\dagger]_{pr}\left(1+\ln\left(\frac{\mu ^2}{M_{\Pi_7}^2}\right)\right) \right. \nonumber\\
   &\left. -\frac{(y_{\Pi _7}^{eq})^*_{j p} (y_{\Pi_7}^{lu})_{ir}}{2M_{\Pi_7}^2}[Y_u^{\dagger}]_{pr}  \left(2 \lambda _{\Pi _7}+\lambda _{\phi } \left(3+2 \ln \left(\frac{\mu ^2}{M_{\Pi_7}^2}\right)\right)\right) \right],\\
    \left[\cC_{eW}\right]_{ij}=&
    -\frac{3g_2}{16\pi^2}\left(\frac{(y_{\omega_1}^{eu})_{jp} (y_{\omega_1}^{ql})^*_{ri}}{16 M_{\omega_1}^2} [Y_u^{\dagger}]_{rp}\left[3 + 2 \ln \left(\frac{\mu^2}{M_{\omega_1}^2}\right)
    \right]\right.\nonumber \\
    &\left.-\frac{(y_{\Pi_7}^{eq})^*_{jp}
    (y_{\Pi_7}^{lu})_{ir}}{16M_{\Pi_7}^2} [Y_u^{\dagger}]_{pr} \left[ 1+2\ln \left(\frac{\mu^2}{M_{\Pi_7}^2}\right)
    \right]\right),\\ 
    \left[\cC_{eB}\right]_{ij}=&\frac{g_1}{16\pi^2} \left(\frac{(y_{\omega_1}^{eu})_{jp}
    (y_{\omega_1}^{ql})^*_{ri}}{16M_{\omega_1}^2}[Y_u^{\dagger}]_{rp} \left[19+10 \ln \left(\frac{\mu^2}{M_{\omega_1}^2}\right) 
    \right] \right. \nonumber\\
    &\left. -\frac{(y_{\Pi_7}^{eq})^*_{jp} 
    (y_{\Pi_7}^{lu})_{ir}}{16M_{\Pi_7}^2} [Y_u^{\dagger}]_{pr} \left[1 + 10 \ln \left(\frac{\mu^2}{M_{\Pi_7}^2}\right)
    \right]\right),\\
    \cC_{HD} =& \frac{-g_1^4}{2880 \pi^2 M^2_{\omega_1}}-\frac{49g_1^4+90(\sqrt{2}\lambda_{\Pi_7}-\eta_{\Pi_7})^2}{5760 \pi^2 M^2_{\Pi_7}}\\
    \cC_{H\Box} =& \frac{(-g_1^4-180\lambda_{\omega_1}^2)}{2880 \pi^2 M^2_{\omega_1}}-\frac{49g_1^4+90(\sqrt{2}\lambda_{\Pi_7}-\eta_{\Pi_7})^2}{5760 \pi^2 M^2_{\Pi_7}},
\end{align}
where we have neglected pieces proportional to SM lepton Yukawas, and leptoquark couplings are defined as in Eq.~\eqref{eqn:LQ_lagrangian}. These expressions were calculated using the \texttt{SOLD} ~\cite{Guedes:2023azv,Guedes:2024vuf} package and are consistent with those in Refs.~\cite{Gherardi:2020det, Aebischer:2021uvt, Gargalionis:2024jaw}.

\subsection{Expressions for $\Delta \kappa_e$ and $\Delta a_e$}
\label{app:lq:expressions}
We give here the expressions for $\Delta a_e$ and $\Delta\kappa_e$ in terms of the $\omega_1$ and $\Pi_7$ model parameters, as defined in Eq.~\eqref{eqn:LQ_lagrangian}, at the mass scales $M_{\omega_1/\Pi_7}=2$~TeV and  $M_{\omega_1/\Pi_7}=10$~TeV. We find
\begin{align}
\Delta a_e &= 10^{-11}\left[180(y_{\omega_1}^{eu})_{13}(y_{\omega_1}^{ql})_{31}^*+(3.0+0.35c^{(c)}_T)(y_{\omega_1}^{eu})_{12}(y_{\omega_1}^{ql})_{21}^*\right.\nonumber\\
&\left.-190(y_{\Pi_7}^{eq})^*_{13}(y_{\Pi_7}^{lu})_{13}
- (2.9 + 0.34\, c^{(c)}_T)(y_{\Pi_7}^{eq})^*_{12}(y_{\Pi_7}^{lu})_{12}\right]\Big|_{2\,\text{TeV}},\\
\Delta a_e &= 10^{-11
}\left[12(y_{\omega_1}^{eu})_{13}(y_{\omega_1}^{ql})_{31}^*+(0.14+0.014c^{(c)}_T)(y_{\omega_1}^{eu})_{12}(y_{\omega_1}^{ql})_{21}^*\right.\nonumber\\
&\left.-12(y_{\Pi_7}^{eq})^*_{13}(y_{\Pi_7}^{lu})_{13}
- (0.13 + 0.013\, c^{(c)}_T)(y_{\Pi_7}^{eq})^*_{12}(y_{\Pi_7}^{lu})_{12}\right]\Big|_{10\,\text{TeV}},
\end{align}
where $c^{(c)}_T$ is a non-perturbative parameter defined in \cite{Aebischer:2021uvt}, expected to be of order 1. And
\begin{align}
\Delta \kappa_e &= \left[(290+73\lambda_{\omega_1})(y_{\omega_1}^{eu})_{13}(y_{\omega_1}^{ql})_{31}^*-(0.23-0.27\lambda_{\omega_1}) (y_{\omega_1}^{eu})_{12}(y_{\omega_1}^{ql})_{21}^*\right.\nonumber\\
&\left.+(270+73\lambda_{\Pi_7})(y_{\Pi_7}^{eq})_{13}^*(y_{\Pi_7}^{lu})_{13}+(0.036+0.27\lambda_{\Pi_7})\,(y_{\Pi_7}^{eq})^*_{12}(y_{\Pi_7}^{lu})_{12}\right] \Big|_{2\,\text{TeV}},\\
\Delta\kappa_e &= \left[(18+2.6\lambda_{\omega_1})(y_{\omega_1}^{eu})_{13}(y_{\omega_1}^{ql})_{31}^*-(16-9.5\lambda_{\omega_1})\times10^{-3}(y_{\omega_1}^{eu})_{12}(y_{\omega_1}^{ql})_{21}^*\right.\nonumber\\
&\left.+(16+2.6\lambda_{\Pi_7})(y_{\Pi_7}^{eq})_{13}^*(y_{\Pi_7}^{lu})_{13}+(10+9.5\lambda_{\Pi_7})\times10^{-3}\,(y_{\Pi_7}^{eq})^*_{12}(y_{\Pi_7}^{lu})_{12}\right] \Big|_{10\,\text{TeV}}.
\end{align}

\subsection{Top FCNCs}
\label{sec:topfcncs}

Given the coupling structure for both leptoquarks and the sizeable couplings needed in order to accommodate a large deviation of the electron Yukawa, one expects also a contribution to FCNCs involving the top quark.
At lepton colliders, top FCNC processes can be probed both in single top production, i.e. $e^+e^- \to tq$, or in the decays of the tops produced in the $t\bar t$ run. In the following, we briefly analyse the prospects coming from both processes, in the cases relevant for us, as well as existing bounds.

\subsubsection*{$e^+e^-\to tq$}

Future prospects for this process have been analysed in \cite{Shi:2019epw}, where all bounds on the relevant effective operators are reported for a 240 GeV run of a circular $e^+e^-$ collider.
The generic constraints one expects on four-fermion operators of the form $[\cC]_{11i3}$, $i=1,2$, are of the order of $0.05$ TeV$^{-2}$.
Setting a mass for the leptoquarks of $M=10$ TeV, one finds that, in both cases
\begin{align}
    [y_{\Pi_7}^{ul}]_{21} [y_{\Pi_7}^{qe*}]_{31} < 14 \,, \qquad
    [y_{\omega_1}^{eu}]_{12} [y_{\omega_1}^{ql*}]_{31} < 14 \,.
\end{align}
For $M=2$ TeV instead one finds
\begin{align}
    [y_{\Pi_7}^{ul}]_{21} [y_{\Pi_7}^{qe*}]_{31} < 0.6 \,, \qquad
    [y_{\omega_1}^{eu}]_{12} [y_{\omega_1}^{ql*}]_{31} < 0.6 \,.
\end{align}

\subsubsection*{$t\to c X$}

The decays to a charm quark and a Higgs or photon necessarily go through chirally-flipping structures ($\cC_{uH}$ or dipole operators), which are suppressed by the electron Yukawa in our setup.
The decay $t\to c Z$, on the other hand, can be mediated by vector structures proportional to the same coupling squared (with different flavour indices), e.g. $[y_{\omega_1}^{eu}]_{13}[y_{\omega_1}^{eu}]_{12}^*$.
In SMEFT, the branching ratio can be written as \cite{Drobnak:2010by}:
\begin{align}
    \mathcal{B}(t\to c Z) = \frac{1}{\Gamma_t} \frac{m_t^3v^2}{32 \pi} \left(1-\frac{m_Z^2}{m_t^2}\right)^2 \left(1 +\frac{2m_Z^2}{m_t^2}\right) \left(|[\cC_{Hq}^{(1-3)}]_{23}|^2 + |[\cC_{Hu}]_{23}|^2\right) \,,
\end{align}
where $\cC_{Hq}^{(1-3)}\equiv \cC_{Hq}^{(1)}-\cC_{Hq}^{(3)}$. Keeping only the log term in the coefficients, one finds
\begin{align}
    [\cC_{Hq}^{(1-3)}]_{23} (m_t) &= \frac{1}{16\pi^2M_{\omega_1}^2} \log\frac{m_t}{M_{\omega_1}} \frac{g_2^2-g_1^2}{6} [y_{\omega_1}^{ql}]_{21}[y_{\omega_1}^{ql}]_{31}^*\\
    [\cC_{Hu}]_{23} (m_t) &= -\frac{1}{16\pi^2M_{\omega_1}^2} \log\frac{m_t}{M_{\omega_1}} \frac{g_1^2}{3} [y_{\omega_1}^{eu}]_{13}[y_{\omega_1}^{eu}]_{12}^*
\end{align}
for the case of the $\omega_1$, and
\begin{align}
    [\cC_{Hq}^{(1-3)}]_{23} (m_t) &= \frac{1}{16\pi^2M_{\Pi_7}^2} \log\frac{m_t}{M_{\Pi_7}} \frac{g_1^2}{3} [y_{\Pi_7}^{qe}]_{21}[y_{\Pi_7}^{qe}]_{31}^*\\
    [\cC_{Hu}]_{23} (m_t) &= \frac{1}{16\pi^2M_{\Pi_7}^2} \log\frac{m_t}{M_{\Pi_7}} \frac{g_1^2}{3} [y_{\Pi_7}^{ul}]_{31}[y_{\Pi_7}^{ul}]_{21}^*
\end{align}
for $\Pi_7$.  Given the current experimental constraint of $\mathcal{B}(t\to cZ) < 1.2 \times 10^{-4}$ (95\% C.L.) \cite{ATLAS:2023qzr}, this doesn't give any meaningful constraint on the couplings.
The FCC-ee projection for $\mathcal{B}(t\to qZ)$ is at  $8\times 10^{-6}$ (95\% C.L.) \cite{khanpour_2023_036fq-ht215}, yielding the still very weak bound
\begin{align}
    y_{12}y_{13} \lesssim 6\times 10^{4},
\end{align}
for a mass $M = 10$ TeV.

\subsection{Plots for $\Pi_7$ leptoquark}
\begin{figure}
    \centering
    \includegraphics[width=1\linewidth]{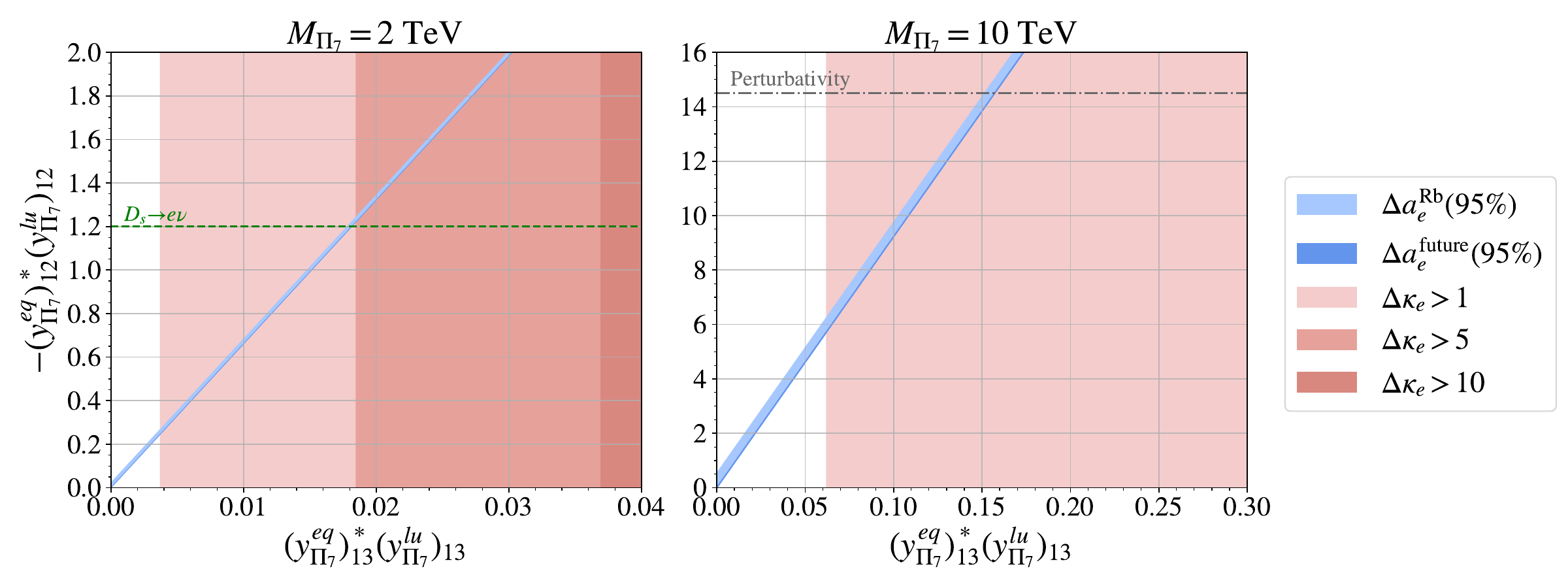}
    \caption{Parameter space for $\Pi_7$ leptoquarks with both electron-top and electron-charm couplings assuming $\lambda_{\Pi_7} = 0$. Red and blue shaded regions are compatible with the stated $\Delta\kappa_e$ enhancement or measurement of $\Delta a_e$ respectively. The green dashed line shows the current 95\% limit on the electron-charm couplings derived from current constraints on $\text{BR}(D^+_s\to e^+\nu)$.}
    \label{fig:pi_7}
\end{figure}

We show plots of the parameter space for the $\Pi_7$ leptoquark, assuming only electron-top and electron-charm couplings, in Fig.~\ref{fig:pi_7}.
\label{app:pi7_plots}

\bibliographystyle{JHEP}
\bibliography{refs}

\end{document}